\documentclass[fleqn,usenatbib]{mnras}

\usepackage{newtxtext,newtxmath}

\usepackage[T1]{fontenc}

\DeclareRobustCommand{\VAN}[3]{#2}
\let\VANthebibliography\thebibliography
\def\thebibliography{\DeclareRobustCommand{\VAN}[3]{##3}\VANthebibliography}

\usepackage{graphicx}	
\usepackage{amsmath}	
\usepackage{tabularx}
\usepackage[table,dvipsnames]{xcolor}
\usepackage[normal]{threeparttable}
\usepackage[nolist]{acronym}
\usepackage{multirow}
\usepackage{subfig}
\usepackage{cleveref}
\usepackage{hyperref}
\usepackage{xspace}

\crefformat{section}{\S#2#1#3}
\crefformat{subsection}{\S#2#1#3}
\crefformat{subsubsection}{\S#2#1#3}

\newcommand{\simba}[0]{\mbox{\fontfamily{lmtt}\selectfont\textbf{Simba}}\xspace}
\newcommand{\moxha}[0]{\mbox{\fontfamily{lmtt}\selectfont\textbf{MOXHA}}\xspace}

\newcommand{\caesar}[0]{{\fontfamily{lmtt}\selectfont \textbf{Caesar}}\xspace}

\newcommand{\GIZMO}[0]{{\fontfamily{lmtt}\selectfont \textbf{Gizmo}}\xspace}

\newcommand{\SOXS}[0]{{\fontfamily{lmtt}\selectfont \textbf{SOXS}}\xspace}
\newcommand{\PyXSIM}[0]{{\fontfamily{lmtt}\selectfont \textbf{PyXSIM}}\xspace}

\newcommand{\emcee}[0]{{\fontfamily{lmtt}\selectfont \textbf{emcee}}\xspace}
\newcommand{\yT}[0]{{\fontfamily{lmtt}\selectfont \textbf{yT\xspace}}\xspace}
\newcommand{\typefont}[1]{{\fontfamily{lmtt}\selectfont \textbf{#1}}}
\newacro{icm}[ICM]{intracluster medium}
\newacro{igrm}[IGrM]{intragroup medium}
\newacro{ism}[ISM]{interstellar medium}
\newacro{igm}[IGM]{intergalactic medium}
\newacro{agn}[AGN]{active galactic nucleus}
\newacro{smbh}[SMBH]{supermassive black hole}
\newacro{bcg}[BCG]{brightest cluster galaxy}
\newacro{bgg}[BGG]{brightest group galaxy}
\newacro{cgm}[CGM]{circumgalactic medium}
\newacro{gmc}[GMC]{giant molecular cloud}
\newacro{lem}[LEM]{Light Elements Mapper}
\newacro{cmb}[CMB]{Cosmic Microwave Background}
\newacro{bhl}[BHL]{Bondi-Hoyle-Lyttleton}
\newacro{grmhd}[GRMHD]{general relativistic}
\newacro{cxb}[CXB]{Cosmic X-ray Background}
\newacro{cmb}[CMB]{Cosmic Microwave Background}
\usepackage[dvipsnames]{xcolor}

\newcommand{\commentblock}[1]{}

\title[Mock X-rays in Simba]{Halo Scaling Relations and Hydrostatic Mass Bias in the \simba Simulation From Realistic Mock X-ray Catalogues}

\author[Jennings \& Davé]{
Fred Jennings,$^{1}$\thanks{E-mail: \href{mailto:Fred.Jennings@ed.ac.uk}{Fred.Jennings@ed.ac.uk}}
Romeel Davé,$^{1,2,3}$
\\
$^{1}$Institute for Astronomy, University of Edinburgh, Royal Observatory,  Blackford Hill, Edinburgh, EH9 3HJ,  United Kingdom \\
$^{2}$University of the Western Cape, Bellville, Cape Town 7535, South Africa \\
$^{3}$South African Astronomical Observatories, Observatory, Cape Town 7925, South Africa
}

\date{Accepted XXX. Received YYY; in original form ZZZ}

\pubyear{2023}

\begin{document}
\newacro{icm}[ICM]{intracluster medium}
\newacro{agn}[AGN]{active galactic nucleus}
\newacro{smbh}[SMBH]{supermassive black hole}
\newacro{bcg}[BCG]{brightest cluster galaxy}
\newacro{cgm}[CGM]{circumgalactic medium}
\newacro{gmc}[GMC]{giant molecular cloud}
\newacro{lem}[LEM]{Light Elements Mapper}
\newacro{cmb}[CMB]{Cosmic Microwave Background}
\newacro{mfm}[MFM]{Meshless Finite Mass}
\newacro{fof}[FoF]{Friends-of-Friends}
\label{firstpage}
\pagerange{\pageref{firstpage}--\pageref{lastpage}}
\maketitle

\begin{abstract}
We present a new end-to-end pipeline for \textbf{M}ock \textbf{O}bservations of \textbf{X}-ray \textbf{H}alos and \textbf{A}nalysis (\moxha) for hydrodynamic simulations of massive halos, and use it to investigate X-ray scaling relations and hydrostatic mass bias in the \simba\ cosmological hydrodynamic simulation for halos with $M_{500}\sim 10^{13-15}M_\odot$. \moxha\ ties together existing \yT-based software packages and adds new functionality to provide an end-to-end pipeline for generating mock X-ray halo data from large-scale or zoom simulation boxes. We compare \moxha-derived halo properties in \simba\ to their emission-weighted counterparts, and forecast the systematic mass bias in mock {\it Athena} observations. Overall, we find inferred hydrostatic masses are biased low compared to true \simba\ values. For simple mass-weighting, we find $b_\text{MW} = 0.15^{+0.15}_{-0.14}$ ($16-84\%$ range), while emission-weighting increases this to $b_\text{LW}=0.30^{+0.19}_{-0.10}$. The larger bias versus mass-weighted values we attribute to the spectroscopic and emission-weighted temperatures being biased systematically lower than mass-weighted temperatures. The full \moxha\ pipeline recovers the emission-weighted hydrostatic masses at $R_{500}$ reasonably well, yielding $b_\text{X}=0.33^{+0.28}_{-0.34}$. \moxha-derived halo X-ray scalings are in very good agreement with observed scaling relations, with the inclusion of lower-mass groups significantly steepening the $L_\text{X}-M_{500}$, $M_{500}-T_\text{X}$, and $L_\text{X}-T_\text{X}$ relations. This indicates the strong effect the \simba\ feedback model has on low-mass halos, which strongly evacuates poor groups but still retains enough gas to reproduce observations. We find similar trends for analogous scaling relations measured at $R_{500}$, as expected for halo-wide gas evacuation.
\end{abstract}




\section{Introduction}

Groups and clusters of galaxies are the largest bound structures in the universe. They have radii up to several megaparsecs, masses up to $10^{15}M_\odot$, and contain tens, hundreds or thousands of galaxies in virial equilibrium, with multi-phase gas filling up the complex inter-cluster or inter-group medium between them. The majority of this gas is hot ionized plasma out to the virial radius \citep{Mcnamara_2007, Kravtsov_2012}, with most of the baryons in the cluster ($\geq 90\%$) existing in this state. This makes galaxy clusters strong emitters in the X-ray \citep{Gursky1973, GurskySchwartz1977}. The temperatures in galaxy clusters are typically between $10^7$ and $10^8$ K \citep{McNamaraNulsen2007}, which results in their glowing with X-ray luminosities usually ranging from $L_\text{X} \sim 10^{43}$ to $\gtrsim 10^{45}$ ergs s$^{-1}$. Groups tend to have luminosities lower by up to several orders of magnitude. To a first approximation the gas and galaxies in these systems trace the background distribution of dark matter, which dominates the mass on these scales and aggregates the baryonic component at the center of the potential of its overdensities \citep{NavarroFrenk1995, MooreGhigna1999, GaoNavarro2008}. The hot gas around galaxies exists in approximate hydrostatic equilibrium, and interacts with feedback from the central \ac{bcg} or \ac{bgg} and satellites, notably through hot gas outflows and jets launched by the central \ac{smbh}. Local thermal instabilities can also condense cold gas from the hot plasma at radii up to several tens of kpc, which sinks through the hot fluid and is thought to accrete onto the central object, possibly fuelling the central \ac{agn} \citep{McDonald_2011,McCourt2012, Yang2016, Beckmann2019, Hitesh_2021} and promoting jet activity. Cooling flows of cold gas are also commonly seen, such as in the Perseus cluster \citep{Fabian1994}, and the need for a heating term to prevent catastrophic cooling has highlighted the importance of energy injection from the \ac{agn} powered by the \ac{smbh} in the \ac{bcg}/\ac{bgg} \citep[see][]{Fabian1994,Peterson2001, Peterson_2003,SijackiSpringel2006,Vernaleo2006,Rafferty2006, Brighenti2006, Mcnamara_2007, HudsonMittal2010,McNamara2012, Gaspari2013, Gaspari2015, Talbot2021, Bourne&Sijacki&Ewald2021,Bourne&Sijacki2021}. \ac{agn} feedback \textit{has} been observed directly, traced by the hot gas, in bubbles and cavities blown and evacuated by powerful hot outflows \citet{Boehringer1993,Churazov2000,McNamara&Wise2000,Fabian&Sanders2000, Birzan2004, Randall2011,Liu&Sun2020, Ubertosi&Brunetti2021}. Understanding the role that hos gas plays in these kinds of environment is naturally crucial for building a coherent picture of feedback cycles and galaxy and halo evolution, and almost all of what we know about the hot medium today comes from studying the X-ray emission from these systems.

In this work we place particular emphasis on the study of \simba\ \textit{groups}. Galaxy groups (we follow similar literature and describe any halo with total mass between $10^{13} M_\odot$ and $10^{14} M_\odot$ as a group) are the middle ground between the halos of $\sim L^\star$ galaxies and galaxy clusters. Groups act as the "building blocks" of the larger mass bin, whilst also acting as cosmic ecosystems where individual galaxies themselves evolve and interact with neighbours. Although in terms of mass they may be na\"ively assumed to be scaled-down versions of clusters, in reality this is far from the truth. The \ac{igrm} ($T \sim 10^7$ K) radiates primarily through metal line emission rather than through Bremsstrahlung (which dominates the hotter \ac{icm}), and this increases the probability of forming a cooler phase ($\lesssim 10^5$ K) in the hot fluid \citep{LovisariEttori2021, OppenheimerBabul2021}, promoting a more complex thermal structure in the halo gas. Indeed it is this metal cooling, particularly through the Iron L-shell emission complex of Fe XVII - Fe XXIV at $0.7-1.2$ keV \citep{GastaldelloSimionescu2021}, that increases the brightness of the \ac{igrm} medium above that of a very faint plasma that would be very difficult to detect at the group temperatures of roughly $0.5-2$ keV \citep{EckertGaspari2021}. Feedback processes affect the gas and stellar content in groups differently from clusters as well. Due to their lower masses and shallower potentials, baryons are more effectively moved and removed through outflows due to stellar and \ac{agn} feedback, and can even become unbound from their host halo; something which is mostly not possible in clusters. Indeed, the radius outside which the physical properties of the gas are dominated by gravity is significantly further out from the halo center in groups of galaxies \citep{AngelinelliEttori2022, AngelinelliEttori2023}. Although difficult to measure, the hot gas fraction is found to be lower in groups than in clusters, and only for the most massive clusters does it approach the cosmic baryon fraction \citep{EckertGaspari2021}. Lower thermal pressures and galactic velocities also result in satellite galaxies being processed differently than in clusters upon in-fall \citep{OmanBahe2021}.
Groups have around an order of magnitude larger number density than cluster-scale halos \citep{JenkinsFrenk2001}, making understanding the physics that shapes them over cosmic time an important milestone in understanding cosmological structure formation and evolution. \\

It is because groups are more dominated by baryonic physics that they are such interesting laboratories. The self-similarity of galaxy clusters, where gas is accretion-shocked to the virial temperature of the cluster (a function of its mass) on in-fall and is though to largely remain in this state at least down to the cluster core, is not necessarily satisfied in the group regime. Gas is both cooled through radiative emission, and widely heated over a large volume by feedback from the central \ac{agn} and from supernovae, creating a more complex multi-phase \ac{igrm} than is attained through simple shock-heating on initial in-fall. Achieving a consistent model of the breaking of self-similarity is crucial both for obtaining accurate mass and temperature proxies, but also for understanding the behaviour and energy scales of feedback, how feedback couples to the \ac{igrm} and how its energy is deposited in the diffuse medium. Deviations from hydrostatic equilibrium, which must be assumed in mass calculations in order to relate the (observable) gas properties to the total halo mass, are expected to be larger in groups, thereby incurring a larger \textit{mass bias} than clusters. Understanding this bias in the groups regime is critical for obtaining accurate mass estimates for cosmological number surveys, among other things.

Deviation from purely thermally-supported hydrostatic equilibrium is believed to be the root cause behind positive mass biases (where $b \equiv 1-M_\text{est}/M_\text{true}$). Additional pressure terms can contribute, and have been shown in both simulations and observations to contribute significantly to the mass bias calculated from the gas profiles alone. One of the most significant non-thermal sources of pressure is from from turbulence in the \ac{igrm} or \ac{icm} \citep{BiffiBorgani2016}. Simulations show that subsonic gas motions which have not yet been thermalised can contribute on the order of $15\%$ of the total pressure in systems, with this fraction being largest away from cluster centers \citep{LauKravtsov2009, AngelinelliVazza2020}. Observations have shown that this fraction is on the order of $5-10\%$ \citep{EckertGhirardini2019}. In addition, gas decelerations away from the cluster center can contribute to a positive mass bias, with an effect in un-relaxed clusters comparable to that of bulk and turbulent motions due to mergers \citep{LauNagai2013, NelsonLau2014}.

Groups are fainter in the X-ray than galaxy clusters, and are also more difficult to detect in optical bands due to the smaller relative number of intrinsically bright central galaxies. Nevertheless we do have some interesting observational results from this mass bin. Significant substructure has been observed in galaxy groups, including X-ray cavities in the \ac{igrm} inflated by hot, low-density, relativistic radio jets, shock-fronts from cavities that can heat \ac{igrm} gas through the turbulent-heating action of their buoyant wakes, cold fronts (distinct from shock fronts by having the cooler gas on the \textit{dense} side) attributed to gas sloshing caused by minor mergers, and filaments of H$\alpha$ emission corresponding to cooled gas $\sim 10^4$ K \citep{MachacekJerius2011,DavidO'Sullivan2011,RoedigerKraft2012,RandallNulsen2015,O'SullivanKolokythas2018,GastaldelloSimionescu2021,BrienzaLovisari2022}. In other wavelengths, jets are ubiquitous in the radio \citep{KolokythasO'Sullivan2020,SchellenbergerDavid2021}, and cold molecular gas has been observed around the central \ac{smbh} in several groups with ALMA \citep{DavidLim2014,TemiAmblard2018} and MUSE \citep{OlivaresSalome2022}. These significant substructures present particular challenges when attempting to treat groups self-similarly and in assuming pressure equilibrium.
Despite the lower detectability and known differences from idealised virialized halos, there are studies which have centered on extending the (X-ray) scaling relations into galaxy groups, and future increases in sensitivity from planned future X-ray observatories will augment these greatly. For an excellent summary of group scaling relations, we direct the reader to the review by \cite{LovisariEttori2021}, and references therein. We briefly summarise the main findings below:
\begin{itemize}
    \item Existing data shows either consistency or a slight steepening of the $L_\text{X}-T_\text{X}$ relation of groups compared to clusters \citep{OsmondPonman2004,EckmillerHudson2011,LovisariReiprich2015,ZouMaughan2016}, which when also taking into account the expected flattening of the relation due to metal cooling, shows the larger impact of heating by feedback on groups compared to clusters. This is through an increase in entropy of cold gas in groups, heating the gas whilst simultaneously reducing the density and luminosity. In the cores of groups, this also reduces the overall gas density through preventing in-fall of cold gas. It has been predicted by \citet{MittalHicks2011} that heating via \ac{agn} feedback begins to dominate over cooling in the \ac{icm} when $kT_{vir} \lesssim 2.5$ keV \citep{LovisariEttori2021}.
    \item  The $L_\text{X}-M_{500}$ relation is less well-studied in the groups regime, and suffers the additional problem that a hydrostatic bias in the mass estimate can directly bias the best-fit scaling slope. There has been work replacing the X-ray mass with a weak-lensing estimate in this relation, and fairly good agreement has been found with the X-ray only scaling relations \citep{SerenoUmetsu2020}, in particular in the groups regime \citep{KettulaGiodini2015}. The $L_\text{X}-M_{500}$ relation is found to either be well-fit by a single power law, or to steepen in the groups regime, due to the aforementioned effect of feedback.
    \item  For the $M_{500}-T_\text{X}$ relation, the slope has been reported to be similar between groups and clusters, yet groups appear to have a lower best-fit normalisation than clusters \citep{EckmillerHudson2011,LovisariReiprich2015}, which will steepen the apparent scaling relation at the low mass end. Due to the absence of direct dependence on complex line emission processes, unlike the luminosity relations the $M_{500}-T_\text{X}$ relation is \textit{not} expected to deviate from the analytic self-similar scaling power law in the absence of feedback. 
    \item  The $M-Y$ relation has been found to be fit very well by a single slope \citep{EckmillerHudson2011,LovisariReiprich2015}, but because of the large measurement errors in the groups regime the intrinsic scatter in the $M-Y$ relation is not well-understood, with the scatter thought to be larger for groups than for clusters \citep{EckmillerHudson2011}.
\end{itemize}

Simulations offer a unique probe into the intrinsic scaling relations and hydrostatic mass bias in systems, via ready access to mass or emission-weighted radial profiles and summed quantities with the desired aperture. Group-finding algorithms also offer precise knowledge on the spherical overdensity mass and radius properties of any given halo.
One of the big focuses in simulations has been in trying to constrain estimates on the degree of hydrostatic mass bias we are likely seeing in observations. \citet{BarnesVogelsberger2021} use their \texttt{Mock-X} pipeline to measure the hydrostatic mass bias in clusters in the \texttt{BAHAMAS} and \texttt{MACSIS} simulations, and report a mass bias of $0.11-0.15$ for their sample, although when considering mock X-ray quantities the bias can be as high as $0.3$ for the largest clusters. Using the same pipeline, \citet{PopHernquist2022} find a mass-weighted mass bias for a combined groups and clusters sample of $0.125 \pm 0.003$, and an X-ray mass bias of $0.17 \pm 0.004$, using over $2,500$ halos. Meanwhile, in work on the \texttt{300 project} of galaxy clusters simulated with \texttt{GADGET-X}, \citet{GianfagnaDePetris2021} find mass-weighted mass biases with a median of $\sim 10 \%$ at $R_{500}$. \citet{BennettSijacki2022} used 27 group and cluster zooms in the \texttt{FABLE} simulation suite to find a median mass bias of $13\%$ using weighted profiles, but with a large scatter.

Analyses have also been performed on simulation scaling relations. For instance, using \texttt{IllustrisTNG} \citet{PopHernquist2022b} studied the \texttt{Mock-X} pipeline to investigate halo scaling relations, finding smooth breaks in each of the scaling relations they consider between $\sim 3 \times 10^{13}- 2 \times 10^{14} M_\odot$, which slopes that usually steepen in the groups regime for $L_\text{X}-M_{500}$, $M_{500}-T_\text{X}$, and $M_{gas}-M_{500}$. In \citet{PopHernquist2022} it is shown that the \texttt{IllustrisTNG} $Y_\text{X}-M_{500}$ relation is steepened at the groups end compare to self-similar. Likewise, previous work on \simba\ by \citet{RobsonDave2020} showed that the intrinsic scaling relations are steeper in groups for $L_\text{X}-T_\text{X}$, $M_{500}-T_\text{X}$, and $L_\text{X}-T_\text{X}$. \citet{YangCai2022} studied the $Y_\text{X}-M_{500}$ relation in \simba\ and also found strong intrinsic deviations from self-similarity going into the group regime, significantly stronger than in \texttt{IllustrisTNG}. \citet{HendenPuchwein2019} use the \texttt{FABLE} zoom sample to investigate the scaling relations, and find that at low redshifts the slopes agree with observations and significantly deviate from self-similar. In particular, the $Y_\text{X}-M_{500}$, $L_\text{X}-M_{500}$, and $M_{500}-T_\text{X}$ relations are all steepened by the effect of \ac{agn} activity on groups.

There are biases that one must be aware of that broadly apply to all simulations. One of the main problems is the lack of cool-core systems in simulations, which presents a challenge when investigating both the scaling relations and the thermodynamic properties of the central gas. \citet{PrattCroston2009} find that cool-core systems bias the luminosity scaling relations of observed systems by increasing the scatter and normalisation, with the distance from the whole-sample best-fit apparently a function of the strength of the cool-core. Almost all simulations fail to reproduce observed entropy profiles down to $\lesssim 0.1R_{500}$ \citep{OppenheimerBabul2021}.

\simba\ has a relatively strong feedback prescription that can remove baryons from halos with high efficiency (see study by \citealp{SoriniDave2022}), which makes it particularly interesting for study in the X-ray compared to other cosmological simulations. It better reproduces the hot gas fraction within $R_{500}$ compared to other simulations, at least below $10^{14}M_\odot$, whereas most other modern suites over-predict the observed gas fractions, sitting above the $90\%$ confidence interval of existing observations as presented by \citet{EckertGaspari2021}. This indicates that \simba\ should more closely match observed luminosity relations, where the total amount of gas sitting in the halo potential determines the magnitude of the emission, as well as the $Y_\text{X}-M_{500}$ relation. It should also produce a more accurate prediction for the hydrostatic mass bias (we predict it will be larger than for other simulations with higher halo baryon fractions), which is directly determined by the density of gas observed within the halo. This success in reproducing observed gas fractions without the need for tuning arises from the strong bipolar AGN jet feedback implemented in \simba\, which effectively removes gas even from group potential wells. Strong feedback in \simba\ is able to substantially affect gas properties out to near the virial radius, and results in \simba\ having distinctly lower normalisations in the density and pressure profiles out to $R_{500}$, when normalised to the analytic expectation values, compared to other simulations. This is neatly shown in the comparisons presented in \citet{OppenheimerBabul2021}.
The manner in which the feedback is fuelled is unique in \simba\ as well, with a torque-limited accretion model implemented for cold gas, which better describes the observed accretion rates compared to the standard Bondi treatment \citep{HopkinsQuataert2011}. This is discussed further in \cref{sec:SIMBA}.
Of course, the diverse implementations on \ac{agn} feeding, feedback mechanisms, and strength, mean that matching simulations to real data through mocks is a key stepping stone to understanding the real drivers of galaxy and group evolution. Because feedback has such a large effect on the baryonic properties of the halo environment within $R_{500}$ and even beyond, mocks can be used to refine the models injecting energy into the \ac{icm} and \ac{igrm} in order to better match observations. \\

Observations of galaxy groups similar in the quality to that achieved in clusters are few, with only a handful to draw from \citep[see][and references therein]{EckertGaspari2021}, and naturally selection effects due to the Malmquist bias are likely to be important. Galaxy groups also present other observational challenges that are not so significant for clusters. A major problem is the universal faintness of groups compared to clusters due to the $L_\text{X}-M_{500}$ relation (although emission from metal lines does boost group luminosities), which means that long exposure times are needed which are not always feasible. Faintness of groups that \textit{are} observable leaves many studies no choice but to extrapolate radial profiles out to $R_{500}$ in order to measure virial quantities, introducing additional sources of bias.

Complexities in emission at group-scale temperatures even muddle the analytic expectations of scaling relations, and result in the prediction of deviations from power-law scaling solely due to local gas properties. The $L_\text{X}-M_{500}$ and $L_\text{X}-T_\text{X}$ scaling relations are expected to flatten at low temperatures, even in the absence of feedback and baryonic physics, solely due to the increased contribution of line emission to the total luminosity \citep{LovisariEttori2021}. \\

One main issue with current works investigating the scaling relations and mass bias in simulations is the reliance on mass or emission-weighted values taken from cells or particles, rather than folding the gas properties through a full mock observation pipeline (although more recent work has begun to address this, for instance \citealp{BarnesVogelsberger2021}). This naturally results in better measurements of the \textit{intrinsic} properties of the systems studied, but does not take into account the projection effects, spectral biases, instrumental responses, and backgrounds with which real data on groups and clusters is inextricably and inseparably linked. Thus, additional scatters and biases are overlooked and comparison to observational data is not entirely apples-to-apples.

With the promise of new observational platforms coming online in the near future, there is a need in the field to take advantage of the several excellent new software packages that have been made available for making realistic mock X-rays, link them together, and develop a universal toolkit to analyse the results. This is what we have done with \moxha. With our easy-to-use python package one can take any compatible simulation box data and generate an end-to-end analysis, leveraging the power of \yT \citep{Turk_2010}, \PyXSIM, \texttt{threeML} \citep{VianelloLauer2015}, and our analysis toolkit, straight out of the box. In future updates of \moxha\ we will add such features such as spatial analysis and stacking, as well as more sophisticated deprojection and spectral-fitting routines.

As of now the most recent X-ray observatory launch was of the eROSITA instrument \citep{PredehlAndritschke2021} in 2019, which has already made important contributions to our understanding of galaxy clusters. Existing eROSITA surveys have probed the scaling relationships of clusters and groups out to low masses and luminosities, have found deviations from self-similarity \citep{BaharBulbul2021}, and performed detailed investigations into properties of individual clusters \citep{IljenkarevicReiprich2021, WhelanReiprich2021} and massive clusters \citep{BureninBikmaev2021}. Planned future missions promise to improve the spatial and spectral resolution of observations, and will be able to probe halo properties, such as degree of self-similarity, with unmatched precision.

One of these future observatories promising great advancements in our understanding of the X-ray Universe is \textit{Athena}. Athena is an L-class ESA mission, due to be launched in the late 2030's. It will consist of two instruments, a Wide Field Imager (WFI) and an Integral Field Unit (IFU). We generate our mocks using the response files of the WFI, in order to take advantage of its wide field of view in order to fit the largest halos we consider into a single chip at low redshift. The WFI has a field of view of $40' \times 40'$, high count rate capability, and will have good spectral resolution over $0.2-15$ keV (FWHM $\leq 80$eV at $1$ keV until end of life) \citep{Meidinger2018}. 

X-ray measurements are the main driving force, alongside simulations, in understanding the physics of group and cluster systems, from radial profiles, to understanding \ac{agn} feedback, to host and satellite galaxy formation and evolution. Despite the rich and numerous history of X-ray missions that have already provided a wealth of data and discoveries associated with groups and clusters of galaxies, there is still much to uncover, and new missions are needed to build on the work already done. Realistic mock imaging of simulated clusters is an essential component in understanding the connection between \ac{igrm} and \ac{icm} properties and the interplay between gravitational and baryonic physics.

In this paper we present a pipeline to analyse cosmological simulations in the X-rays specific to a particular instrument and band, providing an end-to-end pipeline for generating the images, fitting the X-ray spectra, and plotting profiles.  We apply this to the \simba\ simulation to examine the mass bias and X-ray scaling relations in mock data.  \hyperref[sec:SIMBA]{\S2} concerns the set-up and properties of the \simba\ cosmological simulation volume. In \hyperref[sec:Generating Mock Observations]{\S3} we outline our methods for generating, processing, and analysing the mock observations. The hydrostatic mass bias obtained with the mocks, as well as with directly-weighted quantities from the simulation box, is presented and discussed in \hyperref[sec:mass_bias]{\S4}. In \hyperref[sec:Self-Similar Scaling]{\S5} we present the X-ray scaling relations obtained from our mocked halos. We conclude with a discussion of \moxha\ and of our results in \hyperref[sec:Discussion]{\S6}.

\section{SIMBA}\label{sec:SIMBA}
To generate mock observations of simulated clusters we make use of the \simba\ suite of cosmological simulations \citep{DaveAngles-Alcazar2019}. \simba\ utilises the \GIZMO code \citep{Hopkins2015, Hopkins2017}, whose \ac{mfm} approach does not require artificial dissipation terms and so suffers little from artificial diffusion errors. This also means that angular momentum is better conserved compared to more traditional SPH methods, so the effects of torques on forming halos are better captured. Shocks and discontinuities in the fluid are also captured more sharply with \ac{mfm}, and the Lagrangian approach minimises the advection errors which plague mesh-based codes.
Gravity is solved in \GIZMO through GADGET-3's hierarchical tree algorithm which utilises FFT methods for quick and efficient force calculations.
\simba\ models a random $100h^{-1}$cMpc$^3$ cosmological volume, with $1024^3$ each of dark matter and gas particles, with mass resolutions $9.6 \times 10^7 M_\odot$ and $1.82 \times 10^7 M_\odot$ respectively. It adopts a standard $\Lambda$CDM cosmology of \cite{PlanckCollaborationAde2016} ($\Omega_\Lambda = 0.7, \Omega_m = 0.3, \Omega_b = 0.048, h = 0.68, \sigma_8 = 0.82, n_s = 0.97$).
Halos are identified on the fly using a 3-D \ac{fof} algorithm with a linking length set to $0.2$ times the mean inter-particle separation.
Bound structures within halos identified within the simulation volume by running the galaxy-finding \caesar analysis package \citep{Thompson2015}, using a 6-D \ac{fof} on cold gas and stars.

Star formation follows a \citet{Schmidt1959} law based on the molecular hydrogen density, where the fraction in molecular phase is computed via the subgrid model of \citet{KrumholzGnedin2011}.
Metals are tracked within gas particles in \simba\, and also enrich supernovae-driven winds. Metals are also able to be locked into dust within the simulation, for which \simba\ includes models for creation via supernovae and AGB stars, as well as destruction through sputtering, X-ray heating, \ac{agn} jets, and by supernovae explosions \citep{LiNarayanan2019}.

Radiative heating and photoionisation is handled using the GRACKLE-3.1 library \citep{SmithBryan2017}, which also models dust heating and cooling, and the UV background after \cite{HaardtMadau2012}. Cooling through metal lines in GRACKLE makes use of the cooling tables of CLOUDY \citep{SmithSigurdsson2008}. Star formation is used to drive galactic outflows, which are implemented in \simba\ via decoupled two-phase winds. Decoupled here means that the winds do not interact hydrodynamically for up to 2\% fraction of the current Hubble time after launch.  $30\%$ of the particles are ejected in the "hot" phase, with the temperature set by the supernova energy minus the kinetic energy of the wind. The mass loading factor $\eta(M_*)$, the ratio of ejected gas mass to star formation, is assumed to follow scalings with stellar mass $M_*$ based on the FIRE simulations~\citep{Angles-AlcazarFaucher-Giguere2017}. \simba\ also introduces metal-loaded winds, with wind particles extracting some metals from nearby particles in a kernel-weighted manner at launch. Wind velocity scaling are taken from \cite{MuratovKeres2017}, with a wind velocity which is a function of the virial radius of the star and the circular velocity of the star, obtained via a scaling based on the baryonic Tully-Fisher relationship.

Black holes in \simba\ use a \ac{bhl} accretion model for hot gas, under the assumption of spherical symmetry. This occurs for hot ($T > 10^5$K), non-\ac{ism} gas, and is computed \cite{Dave&al_2019};
\begin{equation}
    \dot{M}_\text{Bondi} = \varepsilon_m \frac{4 \pi G^2 M^2_\text{BH} \rho}{(v^2 + c_s^2)^{3/2}},
\end{equation}
where $\rho$ is the density of the hot gas within the accretion kernel, $c_s$ is an averaged sound speed of this gas, and $v$ is an average relative velocity compared to the black hole itself. $\varepsilon_m$ is a tunable efficiency factor, set to $0.1$ in \simba . For cold gas, \simba\ employs a somewhat unique torque-limited accretion model~\citep{Angles-AlcazarDave2017}, after \cite{HopkinsQuataert2011}, with the accretion rate given by 
\begin{equation*} 
    \dot{M}_\text{Torque} \approx \varepsilon_T f_d^{5/2} \times \left(\frac{M_{BH}}{10^8M_\odot}\right)^{1/6} \left(\frac{M_{enc}(R_0)}{10^9M_\odot}\right)  \left(\frac{R_0}{100 \text{pc}}\right)^{-3/2} \times 
\end{equation*}
\begin{equation}\label{eqn:torque-limited-accretion}
    \left(1+\frac{f_0}{f_\text{gas}}\right)^{-1} M_\odot \text{ yr}^{-1}.
\end{equation}
$f_d$ is the disk mass fraction which includes stellar and gas components, $M_{enc}(R_0)$ is the total sum of these two components, $f_\text{gas}$ is the gas mass fraction in the disk. $R_0$ is just the distance from the black hole in which $256$ distinct gas elements are enclosed. The disk and spheroidal components of the gas and stars are separated in the simulation via a kinematic decomposition.
 \ac{ism} gas is all taken to be cold, even if its temperature is above the threshold, because this medium is artificially pressurized in order to resolve the Jeans mass \citep{DaveThompson2016}.
 In total the accretion rate, modulated by a radiative efficiency factor $\eta = 0.1$, is $\dot{M}_{BH} = (1-\eta)\times (\dot{M}_\text{Torque} + \dot{M}_\text{Bondi})$, and upper rates on both components of the total accretion rate are imposed based on the Eddington rate for each black hole.
 
 Black hole feedback is implemented in two modes; \textit{kinetic} and \textit{X-ray}. The kinetic mode is implemented through a sub-grid model and is in turn split into two sub-modes, based on the ratio of accretion compared to the Eddington limit.
At high Eddington ratios ($\gtrapprox0.2$) \ac{agn} drive winds on the order of $10^3$ kms$^{-1}$, which consist of molecular and warmly ionized gas. This is the \textit{radiative} mode.  At lower Eddington fractions the feedback instead occurs in a \textit{jet} mode, where hot gas is driven at velocities of order $10^4$ kms$^{-1}$ that is capable of inflating bubbles and cavities, and evacuating gas from the central regions. Both modes of kinetic feedback are bipolar and have their free parameters, such as velocities and temperatures, calibrated against observations of observed outflows in real systems.
 The radiative wind velocity is parameterised by the black hole mass in the following way:
 \begin{equation}
     v_\text{w,EL} = 500\left(1+ \frac{\log M_{BH} - 6}{3}\right) \text{   km s}^{-1},
 \end{equation}
with the gas being ejected at the \ac{ism} temperature. If the Eddington fraction $f_{Edd} < 0.2$ the feedback is switched to jet mode and the velocity is boosted according to;
\begin{equation}
    v_\text{w,jet} = v_\text{w,EL} + 7000 \log (0.2/f_\text{Edd}) \text{  km s}^{-1}
\end{equation}
which is limited at $7000$ kms$^{-1}$ at $f_\text{Edd}<0.02$. These jets also need a black hole with mass $\gtrsim 10^{7.5} M_\odot$ in order to turn on. This stops small but slowly-accreting black holes from generating jets. Gas ejected in jets is automatically raised to the virial temperature of the halo in \simba\, since observationally jets contain mostly synchrotron-emitting plasma.

Additionally, \simba\ includes \textit{X-ray} feedback, roughly representing the radiation pressure from X-rays off the accretion disk. This feedback mode is only switched on when (full-velocity) jet feedback is also turned on, and so is is determined by the mass accretion rate as a fraction of the Eddington limit. There is also a gas mass fraction threshold, such that X-ray heating is only active if $f_{gas} \equiv M_{gas}/M_* < 0.2$. Gas within the accretion kernel is heated, with the energy input scaled by the inverse square of the distance from the black hole, with a Plummer softening at very small radii. The temperature of non-ISM gas is directly increased according to the heating flux at that position, whereas for ICM gas, which would be projected to just cool quickly, only half the energy is added as heat, with the other half being used to give the gas a kick radially outwards. The gas heating rates due to the X-rays is given in \cite{ChoiOstriker2012};
\begin{equation}
    \dot{E} = n^2(S_1 + S_2),
\end{equation}
with $n$ the local proton number density, $S_1$ the power contribution from Compton heating, and $S_2$ the contribution coming from the sum of photoionization heating. The radiation-pressure change in momentum of the kicked gas is straightforwardly calculated by $\dot{p} = \dot{E}'/c$ where $E'$ is half the total energy available.

The free parameters are broadly tuned to match observations of the galaxy stellar mass function evolution and the stellar mass--black hole mass relation, with the inclusion of X-ray feedback primarily to obtain enough fully-quenched galaxies at $z=0$.  \simba\ has been tested against a large array of other galaxy, halo, and intergalactic medium properties, generally yielding good agreement with available observations~\citep[e.g.][]{Dave&al_2019} at a level comparable or better than other similar simulations. Relevant to this work, \simba\ is shown to match group baryon fractions and X-ray scaling relations derived directly from the particles~\citep{RobsonDave2020}, Sunyaev-Zel'dovich properties of groups~\citep{YangCai2022}, and $z=0$ galaxy quenched fractions~\citep{Dave&al_2019}. Additionally, \citet{HabouzitSomerville2022} report that \simba produces the correct number of X-ray bright \ac{agn} at $z \geq 2$, and also the correct "downsizing" of the peak of the number density of bright versus faint \ac{agn} with cosmic time, driven by efficient \ac{agn} feedback \citep{SchirraHabouzit2021}. Furthermore \simba is shown to match observed Eddington fractions at high $z$ (although there are indications that the \ac{agn} over-accrete at $z<2$ \citealp{HabouzitSomerville2022}). These good matches with observation over a broad range of features make \simba\ a viable platform to explore the physical and observable properties of hot gas within massive halos, which is our aim in this work.

\section{Generating Mock Observations}\label{sec:Generating Mock Observations}

Here we describe the \moxha\ pipeline to generate and fit simulated halo X-ray properties that we have developed for this study.  This involves generating X-ray count maps from particles within a selected simulation region to obtain an X-ray luminosity within a given band, deprojecting the 2-D image into 3-D, fitting the resulting spectra with a plasma model to obtain a temperature, fitting a $\beta$ model to obtain electron density and temperature profiles, and then recovering the mass from these profiles assuming hydrostatic equilibrium.  We also discuss noise subtraction, error estimation, and computing luminosities and aperture-weighted quantities with $R_{500}$.  In the following sections we describe each of these steps in more detail, and finally present the mass completeness of the resulting sample for which the fitting process was successful.

\subsection{X-ray images}

The process of fitting X-ray spectra from simulated groups and clusters begins with generating photons from regions within the simulation volume. To do this we use the \PyXSIM package \citep{BiffiDolag2012, BiffiDolag2013,ZuHoneHallman2016}. \PyXSIM takes the 3D particle data from the simulation and at each point generates a sample of photons based on the physical properties of the plasma at that location. \PyXSIM uses an APEC thermal spectral model \citep{SmithBrickhouse2001}, which depends only on the temperature and metallicity of the gas, to generate line and continuum emissivities. We use the metallicity fields in \simba\ to generate the photons, passing He, C, N, O, Ne, Mg, Si, S, Ca, Fe mass fractions to \PyXSIM. \simba produces a group/cluster population which shows a scatter in luminosity-weighted metallicity from  $[\text{Fe/H}]_\odot = 0.5-2$, as calculated and shown in \citet{RobsonDave2020}. The typical group metallicity is $0.2-0.3$ dex below the solar value, and all clusters but one have comparatively low $L_X$-weighted metallicity, with $[\text{Fe/H}]_\odot \lesssim 0.6$. The authors find that though the  $[\text{Fe/H}]_\odot - T_X$ ($L_X$-weighted temperature) relationship generally matches observations, cold/low mass halos generally have too high an Iron metallicity. They also report that the profiles of individual halos in \simba tend to be too flat, with metal abundance too high in the outskirts of systems, especially for lower mass ($M_{500} < 10^{13.5} M_\odot$) groups. We are using the Athena-WFI response, which does not have the spectral resolution to resolve individual lines. Therefore if there is non-negligible degeneracy between the metallicity and another fitting parameter, \simba's high metal content at large radii may alter the shape of the thermodynamic profiles beyond a few tenths of $R_{500}$ differently compared to what one would expect from real groups.

Before calculating the emission from our systems, we excise cold, dense gas from our dataset in order to remove galaxies, which are artificially pressurized in order to resolve the Jeans length. The threshold we set for this is two-fold. We impose both a temperature cut on the particles at $2 \times 10^5$ K, and a particle cut safely above the \simba\ pressurisation density, at $5.21 \times 10^{-26}$ g/cm$^3$. For $\mu_e=1.155$ this corresponds to a threshold of gas being cut if $n_e > 2.7 \times 10^{-2}$ cm$^{-3}$ (in comparison the \simba\ pressurization density is at $n_H = 0.13$ cm$^{-3}$). We also remove decoupled wind particles, since their hydrodynamic properties are not tracked during decoupling. 

The photons are projected along a line of sight and their energies redshifted by the appropriate amount. The photons are attenuated by foreground absorption on their way to the "observer" using the \textit{tbabs} model \citep{MorrisonMcCammon1983} of foreground absorption, and a set of events are finally written to a SIMPUT catalog file. We then use the complementary software package \SOXS \citep{ZuHoneVikhlinin2023} to convolve these photon events with the Athena-WFI instrumental response to produce a simulated observation of the source. This instrumentation step involves the projection of the events from the source onto the detector plane, performing PSF blurring and dithering, and then convolving the event energies with the response matrix of the instrument, to bin the counts into channels. 

We choose to mock Athena in this work because it has several favourable projected capabilities concerning the observations of group-sized halos at relatively low redshift. Firstly, the large single-chip field of view ($\sim$ 20' $\times$ 20') means that essentially all groups are resolved in single-pointing at $z = 0.05$, with only a slightly higher redshift needed for the most massive clusters we observe. The WFI will also boast good spectral (FWHM $\leq 80$eV at $1$ keV until end of life) and spatial (5'' on-axis half energy width) resolutions, and a low instrumental background with an upper limit of $5 \times 10^{-3}$ cts/cm$^2$/s/keV \citep{Meidinger2018}.  Since we aim to probe scaling relations and mass bias down to fairly low-mass groups, using an existing telescope's configuration would rapidly run out of detectable halos that are distant enough to fit within the field of view.  Nonetheless, \moxha\ could in principle be applied to any telescope and detector configuration available in \SOXS.

We used the default Athena-WFI response files which are packaged with \SOXS \footnote{ Ancilliary Response File (ARF); athena\_sixte\_wfi\_wo\_filter\_v20190122.arf \\ Response Matrix File (RMF); athena\_wfi\_sixte\_v20150504.rmf}. Athena is currently undergoing a period of redesign ("New Athena"), and as such the response characteristics of the instrument are not concrete. This work focuses on the properties of the halo population in \simba, rather than the precise instrumental capabilities of Athena, and as such this fluidity does not present a large challenge to our results. We do believe however that our findings are a good representation of an Athena-like instrument.\\

\begin{figure*}
	\includegraphics[width=\textwidth]{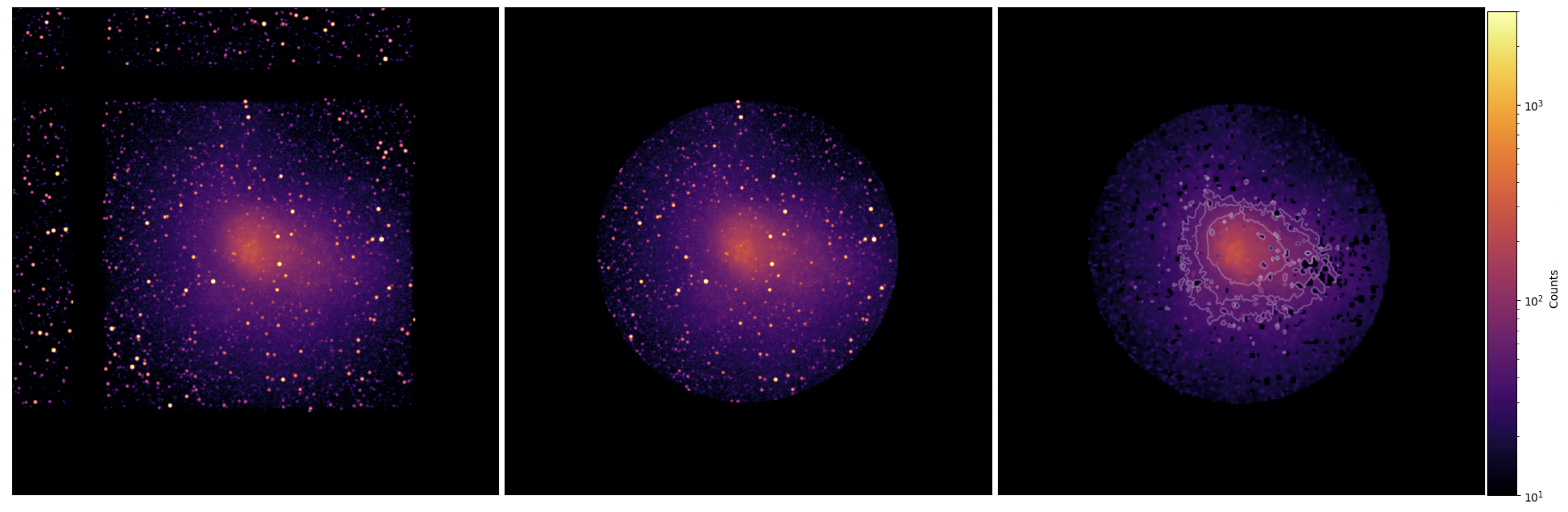}
    \caption{An example of the pre-processing steps the pipeline carries out pre-fitting of the annuli for Athena's WFI is shown. A circle of emission is cut out from the chip to produce the middle panel, and is then cleaned based on a simple background-source masking algorithm to produce a cleaned observation in the final panel. Instrumental background is then subtracted and the emission de-projected.}
    \label{fig:cleaning}
\end{figure*}

\begin{figure*}
	\includegraphics[width=\textwidth]{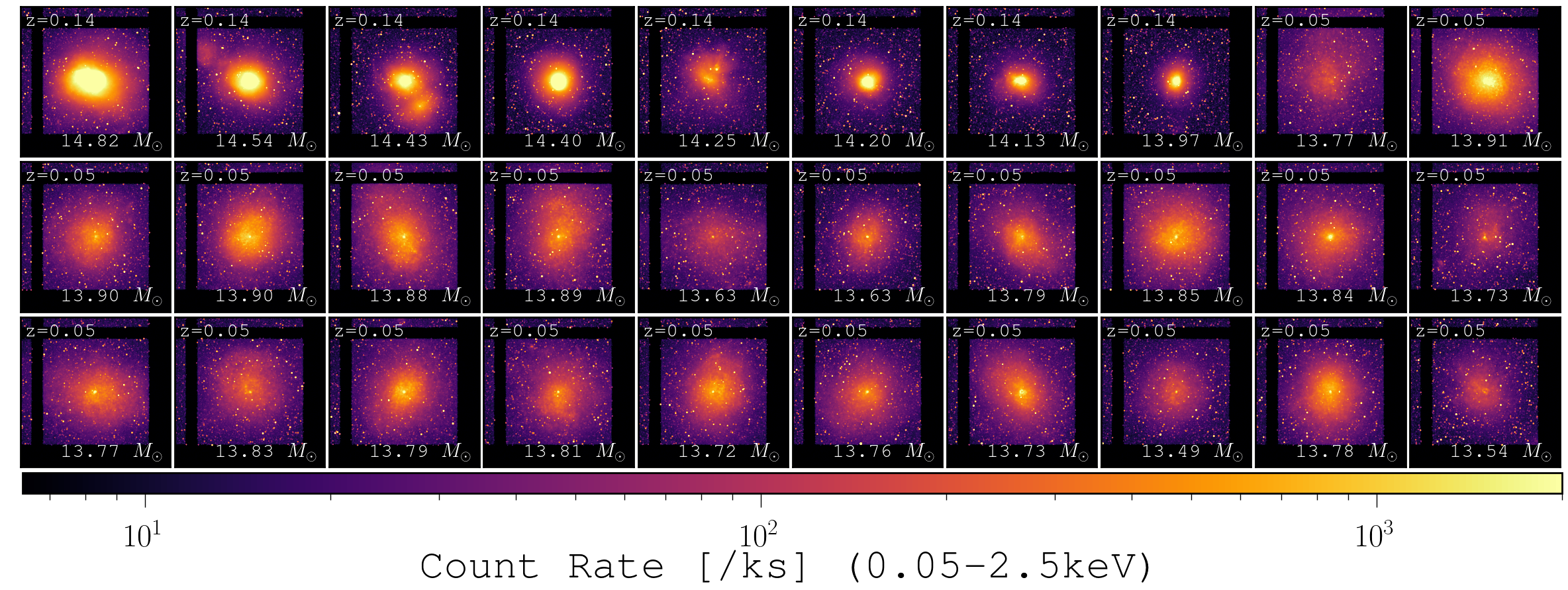}
    \caption{Count rate maps of 30 of the most massive halos found in \simba\ processed through \moxha. Each chip is a single Athena-WFI chip with width $19.06 \ '$.  A wide range of observed morphologies is produced, including regular halos, irregular halos, satellites, and, in some halos, significant visible substructure such as cavities.}
    \label{fig:large_mosaic}
\end{figure*}

We take spectra from successive 2D annuli on the sky centred on the cluster centre as determined by \simba's halo finder. To each annulus we apply a mask though a comparison of counts in the $0.05-2.5$ keV range in each pixel to the median annulus counts. Pixels exceeding twice the median number of counts in the reblocked counts map are masked (this threshold is relaxed to a factor 5 for groups which require a longer exposure, since the emission from the halo itself becomes less uniform at lower halo masses/luminosities),
and a visual inspection of the $2D$ counts maps indicates that this is largely successful in removing obvious point background \ac{cxb} sources (see Fig. \ref{fig:cleaning} as an example). The normalisation of the annulus spectrum is corrected for these subtracted regions by multiplying by the ratio of total to un-masked pixels, though this fraction is generally very close to unity. We deal with unresolved cosmic sources and instrumental noise by subtracting off the counts spectrum of the same annulus region from the background, cleaned using the same counts threshold we used for the signal. We then compute the fraction of channels between $1-2$ keV containing a minimum number of photon counts. If more than  a certain percentage of these channels satisfy this criterion we accept the annulus as having sufficiently good signal-to-noise to be fit. For this work we choose a minimum count of $200$ in $70\%$ of the channels (relaxed to $100$ counts in $70\%$ of channels for long exposure observations/smaller halos). A stricter threshold on annulus acceptance leads to wider annuli which are not so accurately fit by our one-temperature model, whereas a more generous choice leads to thinner annuli which may be ill-fit and more affected by PSF effects.
Although \SOXS does allow one to make mosaics of separate exposures, in this work we focus only on single-pointing observations, and halos for which the spherical overdensity virial radius $R_\text{500,SO}$ does not fit into a single chip at the default redshift we choose to observe the same object at a slightly higher redshift. 
We do not create exposure maps for our observations when generating the annuli, because \SOXS does not currently model spatial differences in effective area across the detector chips. This means that the deprojection will not be affected by different flux characteristics in each annuli. Dither effects may become important near the chip edges, but in practice we rarely measure spectra from very close to the edge of the chip. A subset of the (uncleaned) observations is shown in Fig. \ref{fig:large_mosaic} for illustrative purposes.

\subsection{Deprojection}

X-ray images are 2-D, whereas we are interested in 3-D quantities.
To deproject the measured halo spectra, we assume spherical symmetry and follow the method of direct deprojection as described in \citep{SandersFabian2007}. Starting from the outermost annulus, we calculate a normalised spectrum (counts/energy/volume) for each shell, and then subtract the scaled count from each annulus in which that shell must also be projected through. That is, in annulus $i$ at radius $r_i$ (with the outermost annulus being $i=0)$, we fit the deprojected spectra $S_i$ which is calculated in the following way:
\begin{equation}
    S_i =S_{\text{proj},i} - \sum_{j<i} S_{j,i}  =  S_{\text{proj},i} - \sum_{j<i} \frac{V_{j,i}}{V_{j,j}} \times S_j,
\end{equation}
where $S_{\text{proj},i}$ is the projected spectrum through observed annulus $i$, $S_{j,i}$ is the counts spectrum of shell $j$ observed through annulus $i$, and $V_{j,i}$ is the volume of shell $j$ that is projected through the observed annulus $i$. To calculate the ${V_{i,j}}$, we consider the following; If we observe annulus $i$ with inner and outer $2D$ radii on the sky being $r_{\text{in},i}$ and $r_{\text{out},i}$, then working in cylindrical co-ordinates the volume of shell $j$ for $j<i$ projected through this annulus is
\begin{equation*}
    V_{j,i} =   2 \int ^ {\theta = 2\pi}_0    \int^{r = r_\text{i,out}}_{r_\text{i,in}} \int ^{z = \sqrt{r_\text{j,out}^2 - r^2}} _ {\sqrt{r_\text{j,in}^2 - r^2}} r dr d \theta dz  
\end{equation*}
\begin{equation*}
    = \frac{4 \pi}{3} \bigg[   (r^2_\text{j,in} - r_\text{i,out}^2)^{3/2} - (r^2_\text{j,out} - r_\text{i,out}^2)^{3/2} 
\end{equation*}

\begin{equation}
\hphantom{aaa} - (r^2_\text{j,in} - r_\text{i,in}^2)^{3/2} + (r^2_\text{j,out} - r_\text{i,in}^2)^{3/2} \bigg]
\end{equation}
\begin{equation*}
    V_{j,j} =   2 \int ^ {\theta = 2\pi}_0    \int^{r = r_\text{j,out}}_{r_\text{j,in}} \int ^{z = \sqrt{r_\text{j,out}^2 - r^2}} _ {0} r dr d \theta dz  
\end{equation*}
\begin{equation}
    = \frac{4 \pi}{3} \bigg[   (r^2_\text{j,out} - r_\text{j,in}^2)^{3/2} \bigg]
\end{equation}
Other existing deprojection techniques can cause oscillations between best-fit temperatures, which depend on the annulus sizing and are therefore not physical (see discussion in \citealp{SandersFabian2007}). The method described above has been shown to avoid such oscillations. 

\subsection{Spectroscopic Properties}
Having obtained mock images and spectra of a source we fit the acquired spectra using XSPEC models. XSPEC is an X-ray spectral fitting code developed by \cite{Arnaud1996} which we access through the Python package \texttt{threeML}. We fit a single temperature and electron density (through the normalisation) in the range $0.1-2.0$ keV (observed frame) using a \textit{apec} $\times$ \textit{tbabs} model, with the metallicity free to vary (we use the abundance table of \citealp{AndersGrevesse1989}). The Hydrogen column density for absorption, as well as the redshift, were fixed at the values used in the photon observations. \texttt{threeML} uses the MINUIT minimizer \citep{JamesRoos1975} to maximise the log likelihood of the fit.

Fitting with a single-temperature APEC model results in a best-fit normalisation value $N$. From this we obtain the electron density through
\begin{equation}
    n_e = \sqrt{\frac{ 4 \pi \times 10^{14} \times N \times [D_a \times (1+z)]^2 }{0.82V}},
\end{equation}
where $z$ is the redshift, $D_a$ is the angular diameter distance to the source, and $V$ is the volume of the emitting plasma, which in our case is the projected volume of a shell through a 2D annulus.

We use standard $\beta$-like models for our thermodynamic profiles.
For the temperature profile we adopt the model of \cite{VikhlininKravtsov2006}:
\begin{equation}
    T_{3D}(r) = T_0 t_\text{cool}(r)t(r)
\end{equation}
where outside the central cooling region the profile is dominated by
\begin{equation}
    t(r) = \frac{(r/r_t)^{-a}}{\Big[1 + (r/r_t)^b \Big]^{c/b}}
    \label{eqn:kT_model}
\end{equation}
and the central regions are described by
\begin{equation}
    t_\text{cool}(r) = \frac{(x+ T_\text{min} /T_0)}{x+1} \ \ \ \ \ \ \  \ , \ \ \ \ \ \ \\ x = \Bigg( \frac{r}{r_\text{cool}} \Bigg)^{a_\text{cool}}.
\end{equation}
For the electron density $n_e$ we use the modified $\beta$ model of \cite{VikhlininKravtsov2006};
\begin{equation}
    n_e^2 = n_{e,0}^2 \frac{(r/r_c)^{-\alpha}}{(1+r^2/r_c^2)^{3\beta }} \frac{1}{ \left( 1 + (r/r_s)^\gamma \right) ^\frac{\varepsilon}{\gamma}}.
    \label{eqn:ne_model}
\end{equation}
We find that the fit is good even if we do not include their additional cluster-center term, so we fit all halos with the simpler above model. We fix $\gamma =3, \alpha = 1$ and fit radial profiles obtained through the spectral fitting using \texttt{lmfit} including in the fit any data between $0.1-1.5$ times the $R_\text{500,SO}$ calculated by the halo-finder, which we expect is similar to the X-ray virial radius typically obtained via iteration. We employ a simple "leave one out" method to remove outliers in temperature or density iteratively (if we have sufficient data points) until we are left with a fit that cannot be improved in reduced chi-square by $25\%$ or more by removing a single data point.
\subsection{Total Mass Profiles}
Having obtained thermodynamic profiles we then use the equation of hydrostatic equilibrium to calculate the mass enclosed at each radii:
\begin{equation}
    M(<r) = - \frac{rkT(r)}{G\mu m_p} \left[ \frac{d \log \rho_g(r)}{d \log r } + \frac{d \log T(r)}{d \log r} \right],
    \label{eq:hydrostatic equilibrium}
\end{equation}
with $\mu$ here the mean particle mass. We ensure that each profile is monotonically increasing through a regularisation term (this can be violated if, for instance, the cluster is disturbed or undergoes a merger and there are large gradients in the thermodynamic profiles, in which case the assumption of hydrostatic equilibrium breaks down. This may however occur solely in a local region of the cluster, which can overall still be well-fit by the analytic solution). We use a joint fit for the temperature and density profiles, and minimise the combination;
\begin{equation}
    \lambda \left[\sum_i \left(\frac{\log T_m - \log T_i}{\sigma_{\log T_i}}\right)^2 + \sum_i \left(\frac{\log n_{e,m} - \log n_{e,i}}{\sigma_{\log n_{e,i}}}\right)^2 \right] 
\end{equation}

where we sum over all annuli $i$. $\sigma_{\log n_{e,i}} =  \frac{\sigma_{n_{e,i}}}{ln10 \times n_{e,i}}$ where $\sigma_{n_{e,i}}$ is the (max(plus/minus)) $1-\sigma$ uncertainty as calculated by \texttt{threeML}, and similarly for $\sigma_{\log n_{e,i}}$. $\lambda$ is a regularisation factor which constrains both the fitted temperature profile and also the total mass profile. We have 
\begin{equation*}
    \lambda \ \ \ = \ \ \  D \ \ \  \ \ \ \   \mathrm{ if} \ \ \ \exists  \ \ \ M(<r_{i+1}) < M(<r) \ \ \  
\end{equation*}
\begin{equation*}
   \ \ \ \ \ \ \ \ \ \ \ \ \ \ \ \ \ \ \ \ \ \ \     \mathrm{ or} \ \ \ \exists  \ \ \ \frac{\partial T(r)}{\partial r} \Bigg |_{r>R} > 0 \  , \ \ \ R = 0.8 \ R_\text{500,SO}
\end{equation*}

\begin{equation*}
    \lambda \ \ \ = \ \ \ 1 \ \ \  \ \ \ \ \  \mathrm{ otherwise}
\end{equation*}
where we use profiles generated from the current model fit parameters to generate $M(<r)$ and $kT_\text{X} (\theta_i)$. We select $D$ arbitrarily large. This regularization prevents un-physical mass profiles and also attempts to kill spurious fluctuations in the temperature profiles near the edges of the observation, where in a small number of cases the un-regularized best-fit has an increasing temperature near the virial radius. These cases should be prevented as we evidently expect the temperature to be decreasing with mass outside of the core. We do not apply this regularisation to the mass/luminosity/emissivity-weighted profile fits since unphysical fitting near the edge is simply eliminated in most if not all cases by the high number of data points available.
In addition we carry out an Orthogonal Distance Regression (ODR) \citep{MR1087109} fitting independently on the temperature and density profiles for each halo, in order to determine the effect of including the radial uncertainty on the hydrostatic mass estimates. We weight by the errors in the temperature and density and now also by the width of the bin used to accumulate the spectrum for the spectral fit. Because these fits for the density and temperature profiles are independent in this case we drop the regularisation condition on the mass profile, but this rarely causes an issue. A small number of ODR fits which did not result in good fits to the data, even after varying the initial values of the fit, were excluded from further consideration.

\subsection{Noise Subtraction}
To quantify the noise in the spectrum of the annuli, we use \SOXS to make an observation including only background counts (in the same chip in which we observe our halos), simulating blank-sky observations free of extended sources (halos). We pass these blank-sky observations through the same post-process cut-out and \ac{cxb} cleaning and routine as we do for the actual observations, using the identical annuli radii and median pixel threshold in order to get an estimate of the noise in each annulus. We then subtract the noise spectrum off of the signal during the post-processing routine, before deprojecting.

\subsection{Mass and Emission-weighted data}
In addition to the full mock observation and post-processing, in parallel we take the "true" thermodynamic and mass profiles straight from the particle data. The thermodynamic profiles were mass-weighted (MW), emissivity-weighted (EW) (erg/s/cm$^{-3}$), and luminosity-weighted (LW) (erg/s), with fields generated using \yT's \texttt{add\_\text{X}ray\_emissivity\_field} function, in the same energy band we use for fitting our X-ray observations. This is to compare the results obtained from our full mocks with other work in the field, in which authors almost exclusively use mass or emission-weighted profiles straight from the box, or at best fit summed emission spectra from cells or particles without including photon propagation and projection effects. 
For our weighted profiles we use the same halo center as is used for the instrument pointing, and use the same filtered particle field that we generate for the X-ray emission to remove artificially hot/pressurized gas. These data are then fit using the density and temperature models we use for the observations (Eqns \ref{eqn:kT_model} and \ref{eqn:ne_model}) and the hydrostatic mass estimated and \textit{weighted} scaling relations calculated.

\subsection{Error Calculation}

We estimate errors in our profiles and scaling relations through use of the Goodman-Weare MCMC implementation in \emcee~ \citep{Foreman-MackeyHogg2013}. For each joint fit of $n_e$ and $kT$ (we do four such fits for each halo, one for the \moxha\ profiles and one each for mass, emissivity, and luminosity-weighting) we run $50$ walkers with a chain of length $110,000$, of which $10,000$ steps are of burn-in. After burn-in we thin the chain to obtain $1000$ parameter value samples. The errors on the fit parameters as estimated from \texttt{lmfit} for the individual annuli are typically very small, and so we estimate the true error on the data with \emcee~ in the process, using a more generous prior. The confidence intervals on the two fitted profiles, $kT(r)$ and $n_e(r)$, were estimated by taking the $16^{th}$ and $84^{th}$ percentiles of $1000$ samples from the Markov chain. The confidence intervals on derived profiles (e.g. $M(<r) = M(r, \theta_{kT}, \theta_{n_e})$ and $S(r) = S(r, \theta_{kT}, \theta_{n_e})$) were derived through taking the $1000$ sampled parameter values for both $kT_\text{X}$ and $n_e$ and calculating the $16^{th}$ and $84^{th}$ percentiles of the derived profiles generated with these sample sets. The 1-$\sigma$ errors on derived virial quantities such as $M_{500}$ were calculated through taking the $16th$ and $84th$ percentiles of $M_{500}$ calculated through the sample best-fit parameters drawn from the Monte-Carlo chain. We calculate $R_{500}$ as the radius at which the mass enclosed ($M_{500}$) satisfies 
\begin{equation}
   \frac{M_{500}}{\frac{4}{3}\pi R^3_{500}} = 500 \times \rho_\text{crit}.
\end{equation}\\

\subsection{Aperture-Weighted Quantities}
As well as recording the value at $R_{500}$ of quantities of interest, we also define weighted quantities within the virial radius. This is to better compare to studies which publish some averaged value within this aperture. The weighted quantities within $R_{500}$ for $kT_\text{X}$ and $S_\text{X}$ are (approximately) emission measure-weighted (as in e.g. \citealp{LovisariSchellenberger2020}) through use of the best-fit thermodynamic models in the following way:
\begin{equation}
    kT_\text{X} = \frac{\int ^{R_\text{500,X}}_{r=0.1R_\text{500,X}} kT_\text{X}(r) n_{e,X}^2(r) r^2 dr}{\int ^{R_\text{500,X}}_{r=0.1R_\text{500,X}} n_{e,X}^2(r) r^2 dr},
\end{equation}
\begin{equation}
    S_\text{X} = \frac{\int ^{R_\text{500,X}}_{r=0.1R_\text{500,X}} S_\text{X}(r) n_{e,X}^2(r) r^2 dr}{\int ^{R_\text{500,X}}_{r=0.1R_\text{500,X}} n_{e,X}^2(r) r^2 dr},
\end{equation}
with errors being calculated by calculating $kT_\text{X}$ and $S_\text{X}$ as above using sample model parameters $\{\theta_{ne,i}, \theta_{kT,i}\}$ sampled using \emcee, integrating between $0.1R_{500_{X,i}}$ to $R_{500_{X,i}}$, and then taking the standard deviation of the resulting sample of weighted temperatures/entropies. Although in principle possible to calculate these quantities analytically from the best-fit model values, we integrate these numerically until the result is well-converged. Mass, emissivity, and luminosity-weighted quantities \textit{weighted} within the respective estimated $R_{500}$ aperture for each halo are similarly defined.
$Y_\text{X} = M_{gas} \times T_\text{X}$ is known to be a robust estimator for cluster total mas, given the biases in the components of $Y_\text{X}$ tend to mostly nullify each other \citep{KravtsovVikhlinin2006}. We calculate $Y_\text{X}$ in the following way. We first estimate $M_{gas}$ through

\begin{equation}
M_{gas} = \mu_e m_p \int^{R_\text{500,X}}_{0.1R_\text{500,X}} n_e(r) \times 4 \pi r^2 dr
\end{equation}
where $\mu_e=1.155$ is the mean molecular weight per electron \citep{BaharBulbul2022}. We then multiply this gas mass by the weighted gas temperature calculated as above. We calculate the errors on $Y_\text{X}$ by calculating the spread on $M_{gas}$ with the \emcee~ sample of electron density profiles (summing the gas mass out to $R_{500}$ calculated with the current sample parameters), then combining with the error on the weighted $kT_\text{X}$ in quadrature. 

\subsection{X-ray Luminosities}
To calculate the luminosity of mocked halos, we use the following procedure; firstly, we cut-out all detected photons within $R_\text{500,X}$ of the center of the pointing, which we always coincide with the center of the halo as determined by our halo-finder. Next, we create exposure maps for this circular region using \SOXS ' \typefont{make\_exposure\_map} function using a linear range of energies between $0.5 - 2$ keV (rest-frame) with a spacing of $0.2$ keV, giving $10$ exposure maps in total. We sum the net flux (in counts/s/cm$^2$) in this region using the \typefont{write\_radial\_profile} function with a single bin, and then multiply by the photon energy used to create the given exposure map, to get an energy flux. The luminosity distance is then used to convert this energy flux to a luminosity. We estimate errors on $L_\text{X}$ by repeating this process for the $1 \sigma$ upper and lower bounds on $R_\text{500,X}$ estimated from the \emcee~ sampling of the profile fits. We account for background luminosity by subtracting a blank-sky flux calculated in the same cutout region at each exposure map energy, however we do not currently correct for foreground absorption. This means that our measured luminosities are in principle a lower estimate on the true halo luminosities, although we expect this difference to be small. Some large halos lack a measurement of $L_\text{X}$ since the angular size corresponding to the X-ray estimated $R_\text{500,X}$ is larger than the detector chip width. We define the true luminosity as the luminosity acquired through summing the luminosity between $0.5-2.0$ keV (corresponding to the rest-frame energy range used for our mocked values) using yT's X-ray luminosity field.

\subsection{Completeness}
\begin{figure}
	\includegraphics[width=\columnwidth]{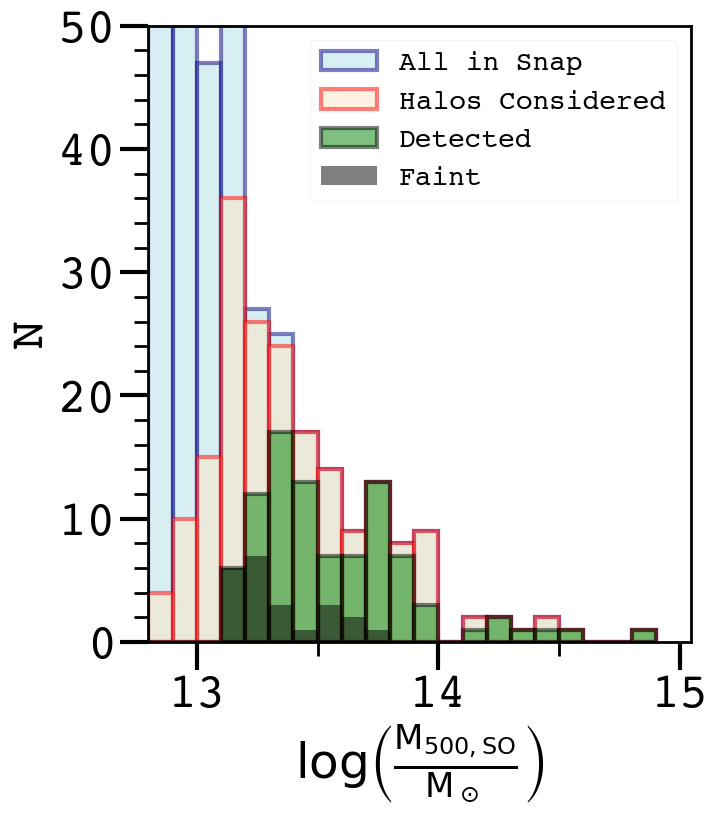}
    \caption{We plot above the completeness of our sample of the $200$ most massive halos identified by \caesar. Note that a range of exposure times were used and this study is not intended as a completeness study. We define "faint" halos to be those that are under-determined in the joint-profile fitting yet still give good profile fits with reasonable error on the hydrostatic mass.}
    \label{fig:completeness}
\end{figure}

We use a range of exposure times at two redshifts ($z=0.051$ for most of our halos and $z=0.138$ for the highest-mass halos) \footnote{The samples at each redshift can be considered independent, because the most massive halos at the higher redshift are still the most massive at the lower redshift, and so are not duplicated in observations.} in order to try to measure as many halos as possible, without investigating the detectability of individual halos with each exposure time. Hence while this is not a rigorous detectability analysis for a particular future survey, we nonetheless plot the completeness of our catalogue in Fig. \ref{fig:completeness} to give an idea for what fraction of halos we obtain good data for in each mass bin. Note that we will only include halos with both good spectroscopic \textit{and} weighted fits in our analysis hereon in, in mass bias and scaling relations analysis, in order to have an apples-to-apples comparison with the fully mocked sample. In this way we will test how well \moxha\ recovers these values in terms of mass bias and self-similar scaling. We consider the 200 most massive halos in these work overall, from $ 12.62 \leq \log M_\text{500,SO}/M_\odot \leq 14.82$, for which we find 92 with good X-ray and weighted profile fits (this is the number of halos in the vast majority of our analyses), with 80 of these also corresponding to good ODR fits and so are used in the ODR estimates for the X-ray mass bias.  In general, the completeness is good down to $2\times 10^{13}M_\odot$, after which it rapidly degrades in detectibility and thus the ability to obtain a good fit. The smallest halo we fit for is at $M_\text{500,SO} = 1.34 \times 10^{13} M_\odot$.

\subsection{Limitations}
While \moxha's X-ray mocking procedure is state-of-the-art, there are still limitations in the approach currently, which will be a focus of further updates to the pipeline. We list some of these here:
\begin{itemize}
\item Point-Spread Effects: Currently we do not account for the instrumental PSF in our treatment of the observed photon maps and spectra. In reality, the observed spectrum is affected by point-spread blurring, and this could impact observed fitted profiles if the annulus width is small enough. We do not believe that our results are significantly affected by PSF effects since the Athena-WFI PSF half-equivalent width (HEW) is planned to be $3-5$" (for comparison, the field-of-view of a single chip is around $1150$"), corresponding to around $2$ pixels \citep{Meidinger2018}, but this is something that will need more detailed consideration in future, especially for studies with \moxha\ of halo centers and for studies with different instruments.

\item Assumption of CIE: Collisional Ionisation Equilibrium is usually assumed in a group and cluster environment, except for in the outskirts of clusters \citep[e.g.][]{WongSarazin2011} and in the Warm Hot Intergalactic Medium, where the densities are low enough that the collisional timescales are comparable to the cluster dynamical timescale. However, it has been shown that rapid changes in plasma conditions, such as those due to merger events or shocks caused by \ac{agn} feedback, can cause this assumption to break down \citep{Prokhorov2010,AkahoriYoshikawa2012}, in which case Non-Equilibrium Ionisation stated must instead be considered. In this work we generated all observations using the assumption of CIE, though in principle it is possible  through \PyXSIM to generate photons under non-equilibrium state conditions. It is necessary in this scenario however to have ionization state fields already calculated and present in the simulation data for every species to be considered.

\item Optical Depth: \PyXSIM currently does not support modelling non-optically thin plasma, so our mock observations do not include any associated effects such as resonant scattering. The \ac{icm} is generally optically thin to most photons but for the strongest resonance lines this is not true \citep{GilfanovSyunyaev1987,BohringerWerner2010}
and resonant scattering has indeed been observed, e.g. in the Perseus Cluster with Hitomi \citep{HitomiCollaborationAharonian2018}. For our studies the effects of increased optical depth are small; they are mainly important where the density is high in the very centers of clusters, and when attempting to recover velocities spectroscopically using emission lines such as the $6.7$ keV Fe XXV K$\alpha$ line \citep{OtaNagai2018}.

\item Milky-Way Foregrounds: In our study we apply galactic foreground absorption of the X-ray photons through the \texttt{tbabs} model of \citet{WilmsAllen2000}, but we do not include an additive hot halo foreground component to the observed spectrum. The Milky-Way has at least two soft-X-ray components in emission with recent \textit{Halosat} \citep{KaaretZajczyk2019} observations pointing towards contributions at $0.18$ and $0.7$ keV from a clumpy MW CGM with a large range of emission measures \citep{BluemKaaret2022}. A more realistic treatment of mock simulated halos would include a model for the MW CGM and account for this with an additional model/s at the spectral-fitting stage.

\item Inverse-Compton Scattering: In the outskirts of clusters where the surface brightness from Bremsstrahlung radiation is typically much lower than in the central regions, inverse-Compton scattering of low-energy \ac{cmb} photons by highly relativistic electrons may become non-negligible \citep{BartelsZandanel2015}. The spectrum of up-scattered photons is determined by the electron population in the scattering region, with relativistic electrons being produced via shocks due to mergers. The magnetic field morphology, alignment and coherence length, and the magnetic field strength, also are needed to determine the spectrum. These are all uncertain quantities in real halos, and are not tracked in \simba, and we do not attempt to model them in this work. It is left to further studies to model this inverse-Compton effect within simulated halos, and compare the fraction of the flux contributed by up-scattered photons to the total spectrum in the outskirts.

\end{itemize}

\section{Hydrostatic Mass Bias}\label{sec:mass_bias}

For many applications particularly in cosmology, it is critical to be able to estimate the halo mass accurately.  It is straightforward to infer this from X-ray profiles assuming hydrostatic equilibrium.  However, this assumption is unlikely to be true in detail, due to for instance disruptions from merging or injection of AGN feedback.  In this section we use our \moxha\ pipeline to test how well the mass estimate from assuming hydrostatic equilibrium and fitting the thermodynamic profiles of the clusters matches the "true" spherical overdensity mass as calculated by our halo-finder, which we denote $M_\text{500,SO}$.  Relative to this, we compute the mass-weighted $M_\text{500,MW}$, emissivity-weighted $M_\text{500,EW}$, luminosity-weighted $M_\text{500,LW}$ and \moxha\ X-ray mass $M_\text{500,X}$. In this way we are able to investigate the bias invoked from transforming mass-weighted to luminosity-related quantities, and quantify exactly what effect the full X-ray mocking has compared to dealing only with emission-weighted quantities, including how instrumental responses as well as foreground absorption, instrumental and \ac{cxb} backgrounds, and projection effects, impact the mass estimate.

\subsection{Mass bias as a function of halo mass}
\begin{figure*}
	\includegraphics[width=\textwidth]{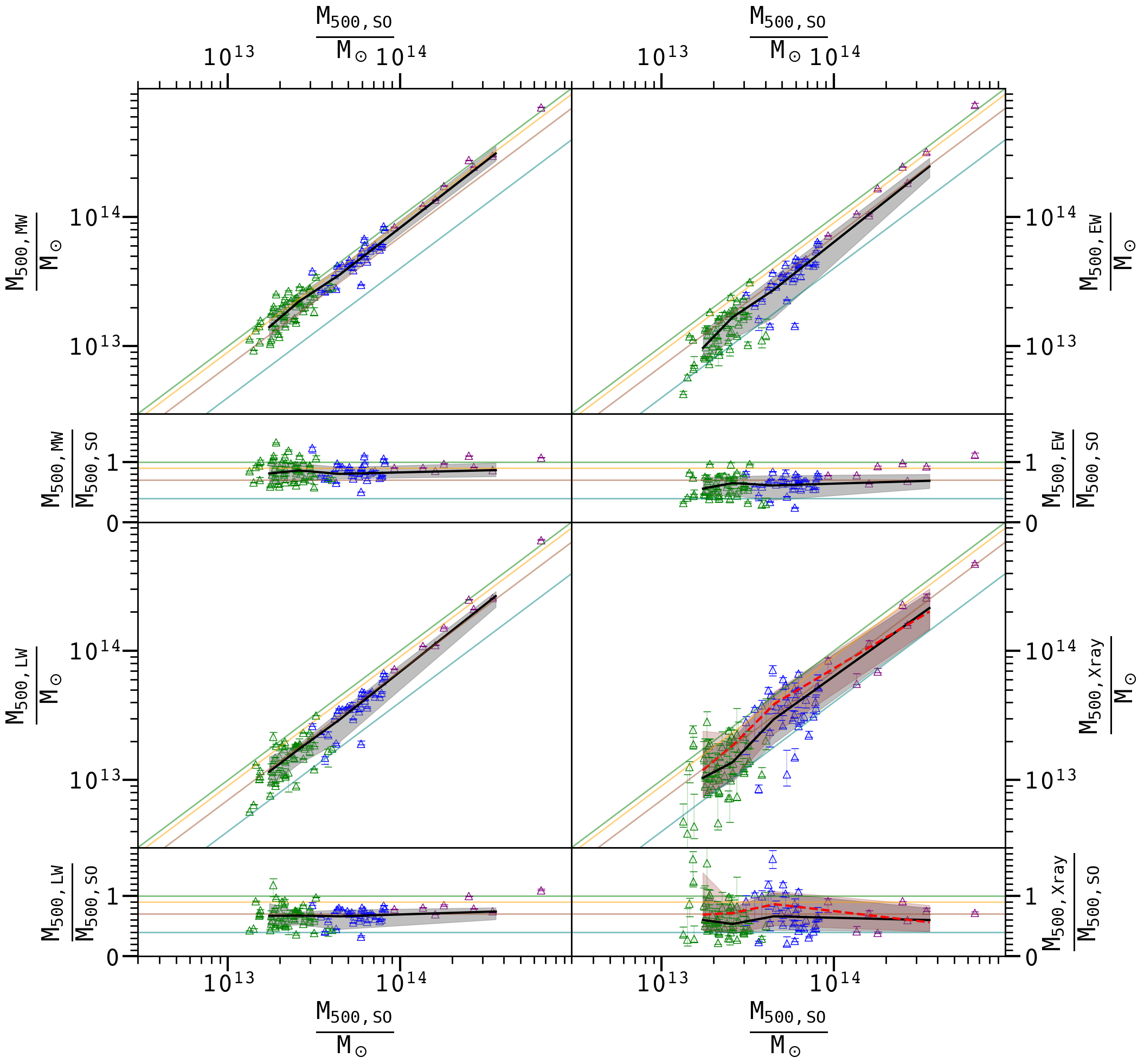}
    \caption{We plot the biases associated with all four of our hydrostatic mass sources, mass-weighted $M_\text{500,MW}$, emissivity-weighted $M_\text{500,EW}$, luminosity-weighted $M_\text{500,LW}$ and \moxha\ X-ray mass $M_\text{500,X}$. The median and $16th-84th$ percentiles are calculated on the residual data and also plotted on both panels, and for the \moxha\ data the median for the ODR fits is plotted as the red dashed line. Biases of $0,0.1,0.3,0.6$ are shown by successive diagonal lines. The mass-weighted bias is the lowest and is comparable to many other studies on simulated mass biases, whereas the emission-related and \moxha\ mass biases are all significantly larger across the whole mass range. The colour scheme is as follows; the purple, blue, and green markers correspond to halos observed at $z=0.138$ for $500$ks, $z=0.051$ for $500$ks, and $z=0.051$ for $5000$ks, respectively.}
    \label{fig:mass_bias_panel}
\end{figure*}

In Figure \ref{fig:mass_bias_panel} we plot the mass-weighted hydrostatic mass $M_\text{500,MW}$ (upper left), luminosity-weighted mass $M_\text{500,LW}$ (lower left), emission-weighted mass $M_\text{500,EW}$ (upper right), and \moxha-estimated mass $M_\text{500,X}$ (lower right) versus $M_\text{500,SO}$.  The smaller sub-panel below each shows the ratio relative to $M_\text{500,SO}$, which are (one minus) the biases. The colour scheme is as follows; the purple, blue, and green markers correspond to halos observed at $z=0.138$ for $500$ks, $z=0.051$ for $500$ks, and $z=0.051$ for $5000$ks, respectively. The black lines represent the running median for the quantity $M_\text{500,HSE}/M_\text{500,SO}$ for the hydrostatic mass estimates obtained using Levenberg–Marquardt least-squares fitted radial profiles. The red dashed line shows the same quantity but for ODR fitted radial profiles. The shaded region shows the $16$th to $84$th percentiles of the mass ratio values with the same color-scheme used. In the large panels the horizontal lines denote biases of $0,0.1,0.3,$ and $0.6$, and these are reproduced as horizontal lines at constant bias in the lower panels.

For most halos, $M_\text{500,SO}$ tends to be higher than any of the weighted masses, as most of the points lie below the 1:1 line (green diagonal).  The typical bias can be seen in the smaller sub-panels. The mass-weighted biases (upper left panel) are the lowest we find. The median bias for the whole sample is $0.15\pm 0.019$, which is in line with the measured mass bias from other simulations, and this bias is fairly constant across the entire mass range. There is however a slight increase in $b$ below $M_\text{500,SO} \approx 4 \times 10^{13} M_\odot$, signalling an inherent mass dependence for the bias.

The luminosity-weighted biases (lower left panel) are similarly distributed, with a near-identical inter-percentile width to the mass-weighted case, but for a shift in the median value. We find the typical bias in this case to be $0.30\pm0.02$, signifying already a large bias invoked just emission-weighting the particles, as well as a slight asymmetry in the distribution of the bias towards a high-bias tail. 

The emissivity-weighted bias (upper right) is broadly similar to the luminosity-weighted one, although emissivity-weighted hydrostatic mass estimates tend to be more biased and the distribution of the bias is flatter. This is due to a stronger dependence on mass than is seen in the LW sample, with a larger EW bias clearly seen at smaller halo masses. The mean emissivity-weighted bias over the whole mass range is $0.36\pm0.02$.

To understand the LW vs. EW biases, consider a uniform-pressure multi-phase fluid shell with $\rho \propto P/T$, and the gas emissivity having approximate dependence on the density and temperature of the form $\varepsilon \propto \rho^2T^{1/2}$, we would then expect the emissivity (from Bremsstrahlung) to satisfy $\varepsilon \propto \rho^{3/2} \propto T^{-3/2}$. Weighting by emissivity will therefore upweight low-$T$ and thus tend to bias the temperature low compared to mass weighting.

For the luminosity-weighted quantities, for a constant simulation particle mass we have that $L \sim \varepsilon \times V \sim \varepsilon/\rho$, so that $L \propto \rho^{1/2} \propto T^{-1/2}$. This explains why the luminosity-weighted mass bias tends to be less than the emissivity-weighted counterpart.

Finally, we examine the spectroscopic X-ray derived masses and bias (lower right) obtained through the full \moxha\ pipeline.
Over the entire mass range, the median bias is further increased relative to any of the other weighted biases, reaching  $0.41\pm0.03$. This spread is comparable to observations, for instance \citet{EttoriGhirardini2019}.
The luminosity-weighted profile has a comparable median across most of the mass range but does not show the increase in bias below $4-6 \times 10^{13} M_\odot$ that we see in $M_\text{500,X}$ and $M_\text{500,EW}$.
 
The \moxha\ trend also has a larger scatter than the emissivity-weighted in the groups regime, and we associate this with having few radial data points for faint groups, although other sources of scatter, such as the effects of (de-)projection, could also be a source. 

We attempt to better account for low bin-number profiles by including the bin-widths in our profile fitting through an ODR treatment. We see a slightly different median from downwards of $M_\text{500,SO} = 6 \times 10^{13} M_\odot$ compared to the default Levenberg–Marquardt least-squares fitting. Above this mass the bins tend to be small enough that no real difference is made to the mass estimates, but below this mass including the bin widths in the fitting tends to reduce the mass bias and make the median bias actually \textit{smaller} in the groups regime compared to clusters. The overall median bias for the ODR fits including all halo masses is $20\%$ smaller relative to the Levenberg–Marquardt value, though the spread is larger, with a value of $b=0.33 \pm 0.04$.

\begin{figure*}
	\includegraphics[width=\textwidth]{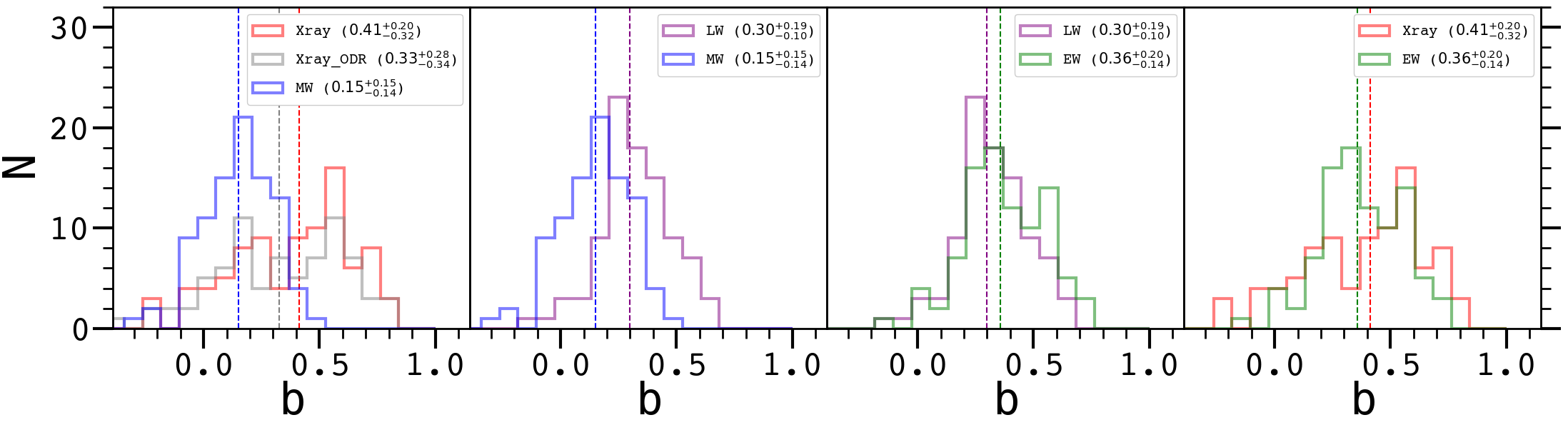}
    \caption{We plot the bias distributions based on different weighting schemes over the entire halo mass range, with medians and $16^{th}$ and $84^{th}$ percentiles quoted as a spread for each histogram. The position of the median is also denoted by a vertical dashed line using the same colour scheme. The mass-weighted bias is around the $15\%$ mark, in line with biases found in similar work. The emission and \moxha\ mass biases are all systematically higher, between $\sim30-40\%$.}
    \label{fig:mass_bias_hists}
\end{figure*}

\begin{figure}
	\includegraphics[width=\columnwidth]{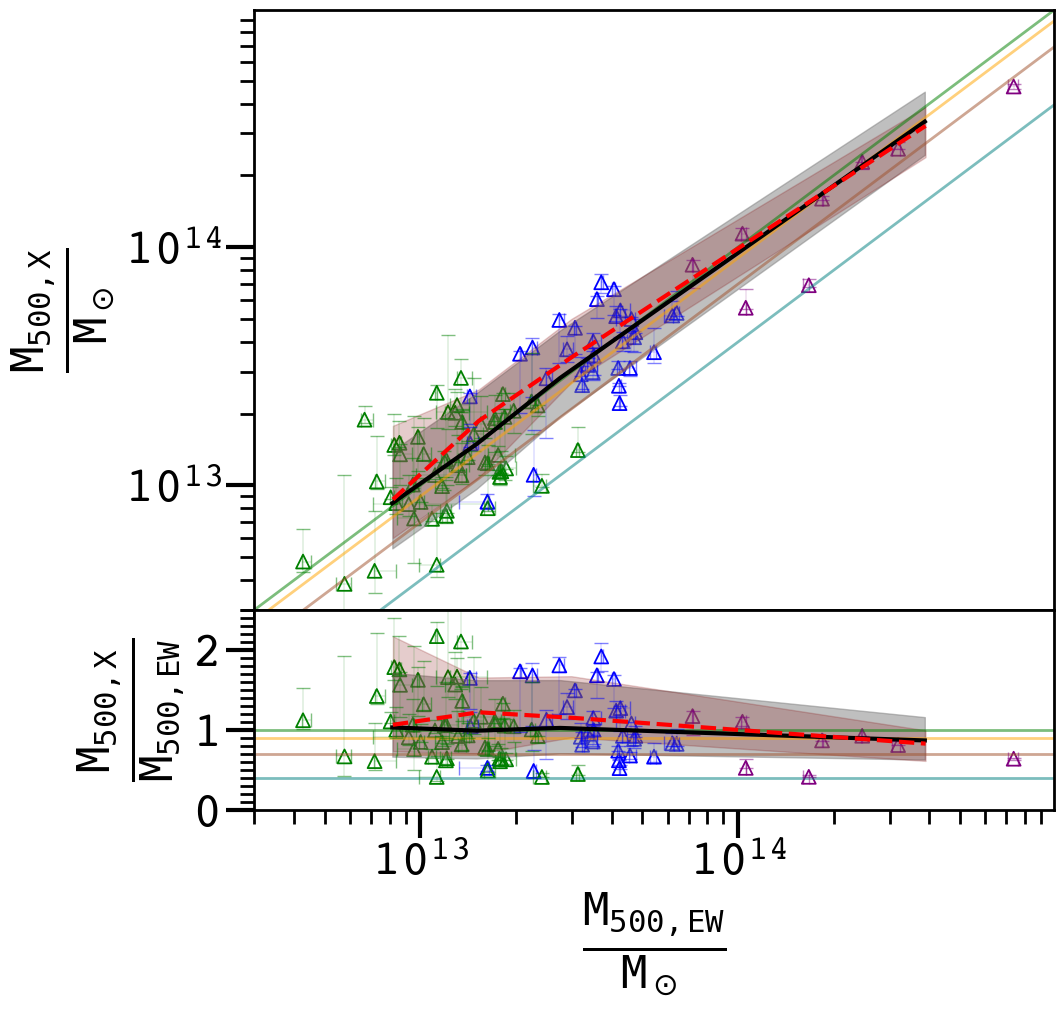}
    \caption{We plot the \moxha\ mass estimates against the emissivity-weighted values explicitly here, for clarity of comparison. The median bias between the two is $\sim 0-0.1$ across the entire mass range, and for the ODR fits (red dashed line) the median bias tends to be marginally lower for the \moxha\ sample compared to EW.}
    \label{fig:Mx_vsM_EW}
\end{figure}

From the profiles, the bias from the true emission/luminosity-weighted temperatures to our X-ray spectroscopic measured temperatures for our most massive clusters is minimal and the uncertainty at each radius tends to be low. This is consistent with the findings of \citet{MazzottaRasia2004} and \citet{RasiaMazzotta2005} , who use simulations of a two-phase \ac{icm} to show that a single-temperature fit is statistically acceptable for high-mass, high-temperature halos (though this may still be biased compared to the "true" temperature). In their case the threshold for this is that the lowest temperature component has $kT > 2$keV. The single-temperature fits we find for these two largest clusters has $kT_\text{X} > 4$keV for all annuli, so although we do not know the temperatures of specific major components, this condition is likely satisfied for the major temperature components which may be present.

The median bias in the \moxha\ values we find using \simba\ is higher than that studied recently in other simulations, notably with \citet{PopHernquist2022} ($0.17 \pm 0.004$ with synthetic X-ray observations in \texttt{IllustrisTNG}), \citet{GianfagnaDePetris2021} ($0.14^{+0.11}_{-0.13}$ with a mass-weighted sample in \texttt{GADGET-X}), \citet{BennettSijacki2022} (a mixed mass/volume weighting scheme used with \texttt{FABLE} data finding $b \approx 0.13$), and higher than the \simba\ mass-weighted result, for which we find a median of $0.15 \pm 0.02$.

Our bias results are in better agreement with the bias needed to ease the tension between expected number counts of $z < 1$ clusters from CMB and tSZ measurements; a bias of $b \approx 0.3-0.4$ or even higher is needed to reconcile the two, with \citet{SalvatiDouspis2018} reporting a bias of $(1-b) = 0.62 \pm 0.07$ required to reconcile the CMB constraints from with tSZ number counts. This is due to the fact that a larger mass bias results in a higher value for $\sigma_8$ which is preferred by CMB primary anisotropies as measured with Planck \citep{PlanckCollaborationAde2014}. The scatter in the mass estimates is known to arise from irregular radial temperature distributions, and also instrumental background \citep{RasiaEttori2006}.

We checked that there is no apparent systematic bias introduced between our samples of different exposure length or redshift. Over the redshift range tested ($0.138-0.051$) no significant halo evolution is expected that would introduce bias into the true values of $M_{500}$ or into the radial structure of the \ac{icm} in general, and indeed there is a smooth transition from this redshift to the sample at lower redshift. Furthermore there does not appear to be systematic differences between different exposure times. However, since we are forced to use longer exposures for smaller halos, which will most likely have more complex thermal structures and are less bright compared to the background, one must be mindful to take this into account when comparing the scatter between these two populations.

In summary, the mass-weighted bias we find in the full-physics \simba box is fully consistent with similar studies carried out with other simulations, with a value of $\sim 15\%$ across the entire mass range. A larger bias is incurred as soon as the weighting is performed on emission-related values, with a median bias of $30\%$ and $36\%$ seen for fits to luminosity-weighted and emissivity-weighted radial profiles respectively. For our \moxha sample which consists of a fully-mocked sample with \textit{spectroscopically}-determined profiles, we find a slightly larger bias at $41\%$. However when we account for the larger radial bins in the groups regime through an orthogonal-distance least-squares regression, the median bias is completely consistent with the luminosity-weighted value, at $33\%$, although there is a larger uncertainty on this value. There is mild mass-evolution in the emissivity-weighted and (non-ODR) \moxha sample mass biases, with a slight increase in median bias below $\sim 5 \times10^{13}M_\odot$ seen for both.

\commentblock{

Xray_HSE_M500 with 92 Samples, Standard Errors on Median: 0.41 +/- 0.0342
Xray_ODR_M500 with 80 Samples, Standard Errors on Median: 0.33 +/- 0.0436
MW_HSE_M500 with 92 Samples,   Standard Errors on Median: 0.15 +/- 0.0190
LW_HSE_M500 with 92 Samples,   Standard Errors on Median: 0.30 +/- 0.0186
EW_HSE_M500 with 92 Samples,   Standard Errors on Median: 0.36 +/- 0.0223

}

\subsection{Bias distributions}

\begin{table*}

\begin{tabular}{||c| cccc||}
\hline
 & EW & LW & \moxha\ (ODR) & \moxha\ \\ \hline\hline
MW & (0.533, 2.70e-12) & (0.478, 6.81e-10) & (0.424, 8.54e-08) & (0.511, 2.69e-11)  \\
\moxha\ & \cellcolor{green} (0.163, 1.74e-01) & (0.272, 2.14e-03) & \cellcolor{green} (0.12, 5.29e-01) & -  \\
\moxha\ (ODR) & (0.261, 3.68e-03) & (0.228, 1.63e-02) & - & -  \\
LW & \cellcolor{green} (0.185, 8.64e-02) & - & - & -  \\
 \hline \hline
\end{tabular}

\caption{We plot the results of performing a two-dimensional K-S test on the sample mass biases arising from different weighting schemes, as well as the \moxha\ X-ray fits. The data in each cell is in the format (\textit{D}, \textit{p-value}), where \textit{D} is the Kolmogorov-Smirnov statistic. Cells which have a p-value such that at the $95\%$ level of significance we cannot reject the null hypothesis that the two samples were drawn from the same distribution are coloured green.}
\label{table:mass_bias_ks_tests}
\end{table*}

In Figure~\ref{fig:mass_bias_hists} we show the distribution of bias values in each weighting scheme. In each panel, we show histograms of $b$ for selected cases for comparison.  Blue lines correspond to mass-weighting, purple to luminosity-weighting, green to emission-weighting, red to our X-ray \moxha\ biases, and grey is spectroscopic biases analyses with the ODR fitting scheme.  Median bias values with $1\sigma$ scatter are indicated in the legends.

The left panel shows the overall bias going from mass-weighting to our \moxha\ pipeline-determined biases. The bias is clearly much higher with \moxha, with a mode at $b>0.5$ versus 0.15 in the mass-weighted case. The scatter is much larger also, showing an asymmetric tail to high bias values.  The ODR fitting tends to give a slightly lower bias, but still qualitatively similar. This indicates that one should not compare mass-weighted bias predictions from simulations to biases inferred from observations.

The remaining three panels show how the bias distribution changes as we change the weighting scheme.  The luminosity-weighted bias shows a shifted distribution relative to mass-weighted, but the scatter is broadly similar. The emission-weighted case is similar.  The full mocks however introduce a significant scatter, as well as further increasing the median bias. 

We test whether the differences between different weighted or \moxha\ mass estimates are statistically significant via the two-sample Kolmogorov-Smirnov (K-S) test \citep{Hodges1958}. 

We tabulate the results from this test in Table \ref{table:mass_bias_ks_tests} for our mass-weighted, \moxha, emissivity-weighted and luminosity-weighted hydrostatic mass estimates. We denote the test between two given mass bias samples A and B as K-S$(b_A,b_B)$.

At the $95\%$ level ($p>0.05$), only the \moxha\ (non-ODR) and emissivity-weighted samples, as well as the luminosity and emissivity-weighted samples, are found to be drawn from the same distribution. The $p$-value of K-S$(b_\text{EW},b_\text{X})$ is $0.17$, and of K-S$(b_\text{EW},b_\text{LW})$ is $0.086$. From Fig. \ref{fig:mass_bias_hists} we see that the \moxha\ and EW distributions are indeed similar with the only real difference being a flatter distribution and a peak at slightly higher bias for the \moxha\ quantities. A direct comparison between the mass estimates for the\moxha\ and EW samples is shown in Fig. \ref{fig:Mx_vsM_EW}, clearly showing the similarity between the two. The strongest evidence of deviation from the null hypothesis comes in the EW versus MW statistic ($p-$value $= 2.7 \times 10^{-12}$), with the \moxha\ and luminosity-weighted values having a similarly strong but slighter larger p-value when compared against the mass-weighted sample. This indicates that using an emission-weighted bias measure could plausibly mimic a full X-ray \moxha\ analysis, though we find the latter to be mildly larger.

\subsection{Bias scatter}

\begin{figure}
	\includegraphics[width=\columnwidth]{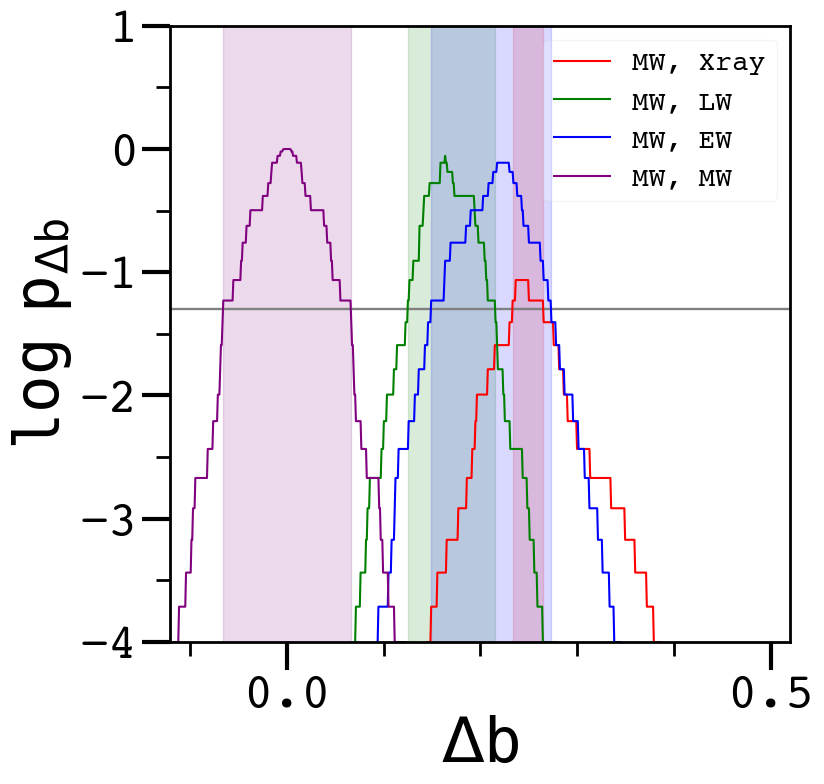}
    \caption{We plot the p-value for the shifted K-S test against the mass-weighted bias distribution.The bias shift value is shown as the independent variable, and the MW-MW relation with $\Delta b$ is shown for reference. The $5\%$ significance level is illustrated by the horizontal line. The shape of the luminosity and emissivity-weighted bias distributions are consistent with them being drawn from the same distribution of the mass-weighted bias with a shifted mean. The \moxha\ distribution is consistent with this conclusion over a smaller $\Delta b$ range and to a lower maximal significance.}
    \label{fig:2d_shifted_ks_MW}
\end{figure}

To test the similarity in the scatter, we also define a ``shifted" K-S statistic where we shift each distribution by $\Delta b$, relative to the mass-weighted bias distribution.  If $\Delta b$ is exactly the difference in median biases, we expect the K-S $p-$value will be maximised at a value that indicates the similarity in the shapes of the two distributions.

Fig. \ref{fig:2d_shifted_ks_MW} shows the resulting $p-$value for each bias sample calculated against the mass-weighted sample, relative to $\Delta b$. The colour scheme is the same as in Fig.~\ref{fig:mass_bias_hists}, and the colour-matches vertical bands indicate the range encompassing $p>0.05$.

We see that the \moxha\ sample has the narrowest range in which the null hypothesis would not be rejected. This range is $(0.233,0.264)$ compared to $(-0.065,0.066)$, $(0.125,0.215)$ and $(0.149,0.272)$ for the mass-weighted, luminosity-weighted and emissivity-weighted samples respectively. The similarity between the significance range and maximal significance between the mass-weighted sample and all but the \moxha\ sample, and especially with the luminosity-weighted sample, indicate that these could well be the same distributions only differentiated by a shifted mean. This is less likely to be the case for the \moxha\ sample, with additional sources of bias introduced through the observation and spectral-fitting procedure.

\section{X-ray Scaling Relations}\label{sec:Self-Similar Scaling}

\begin{figure*}%
    \centering
    \subfloat[\centering Mass-weighted]{{\includegraphics[width=7.2cm]{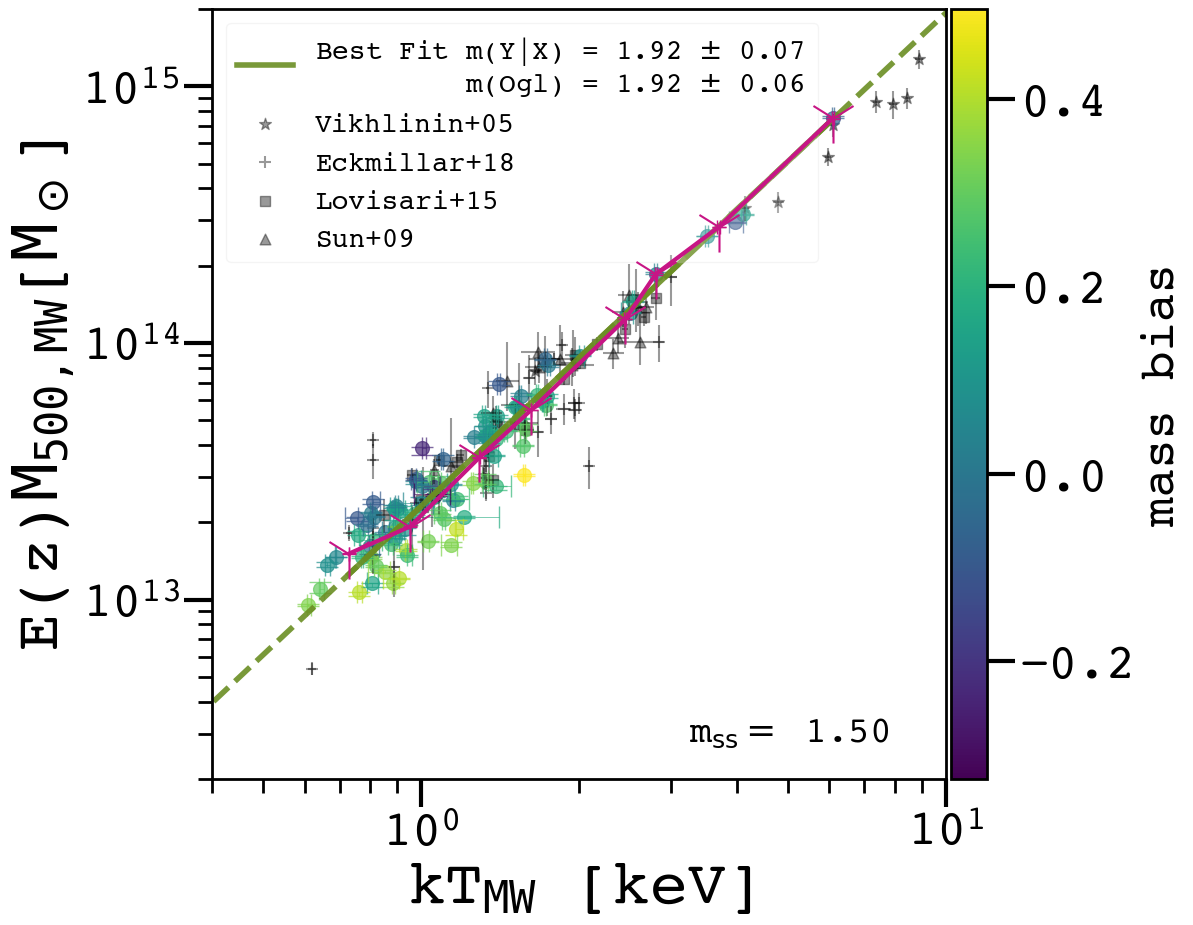} }}%
    \qquad
    \subfloat[\centering \moxha]{{\includegraphics[width=7.2cm]{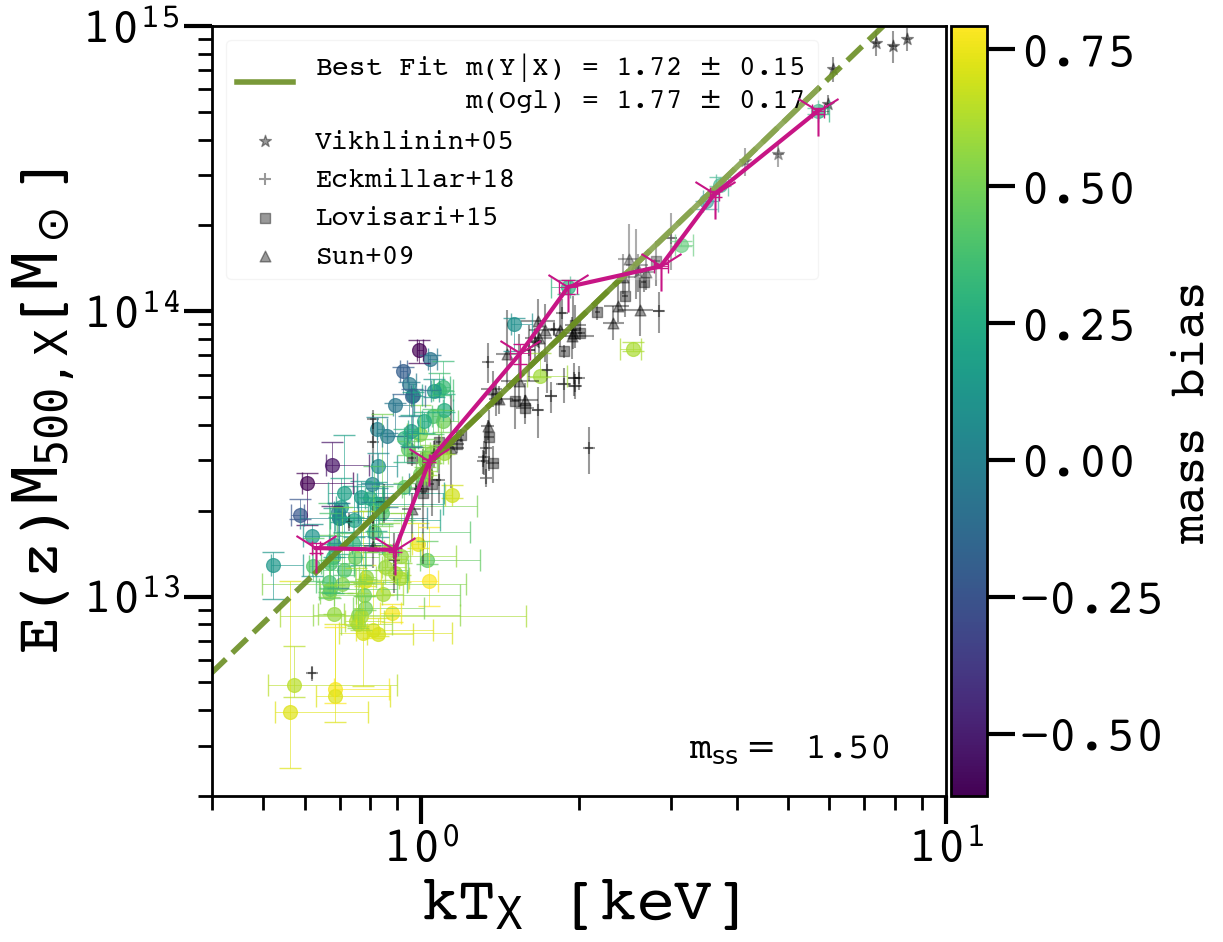} }}%
    \caption{$M_{500}-T_\text{X}$ relation for the mass-weighted (\textit{left}) and fully-mocked \moxha\ (\textit{right}) samples. Here, $T_\text{X}$ is measured as the (core-excised) weighted value out to $R_{500}$, determined from the hydrostatic mass estimate. Both MW and \moxha\ samples agree well with observational data, and both are steeper than self-similar. The \moxha\ sample has increased scatter at the low mass end due to additional mass and temperature bias.}%
    \label{fig:M_vs_kT}%
\end{figure*}

\begin{figure*}%
    \centering
    \subfloat[\centering True Emission]{{\includegraphics[height=6cm]{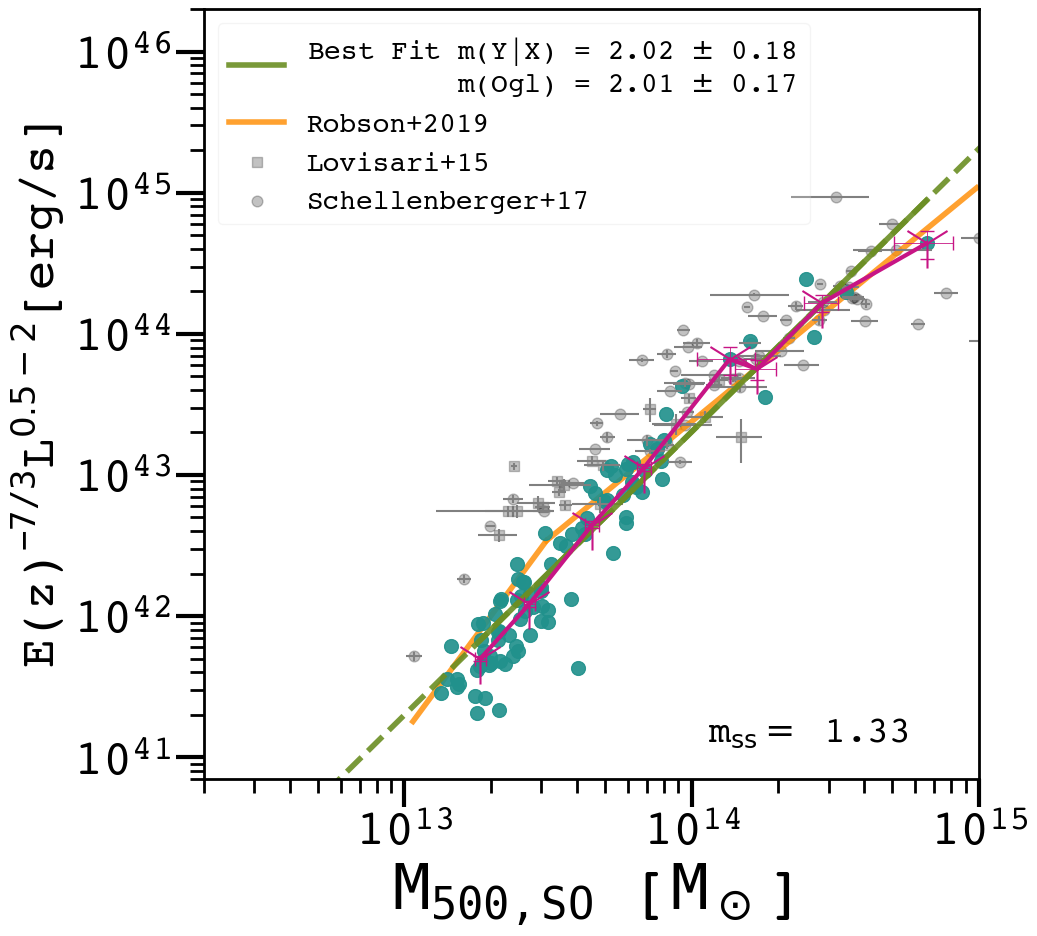} }}%
    \qquad
    \subfloat[\centering \moxha]{{\includegraphics[height=5.8cm]{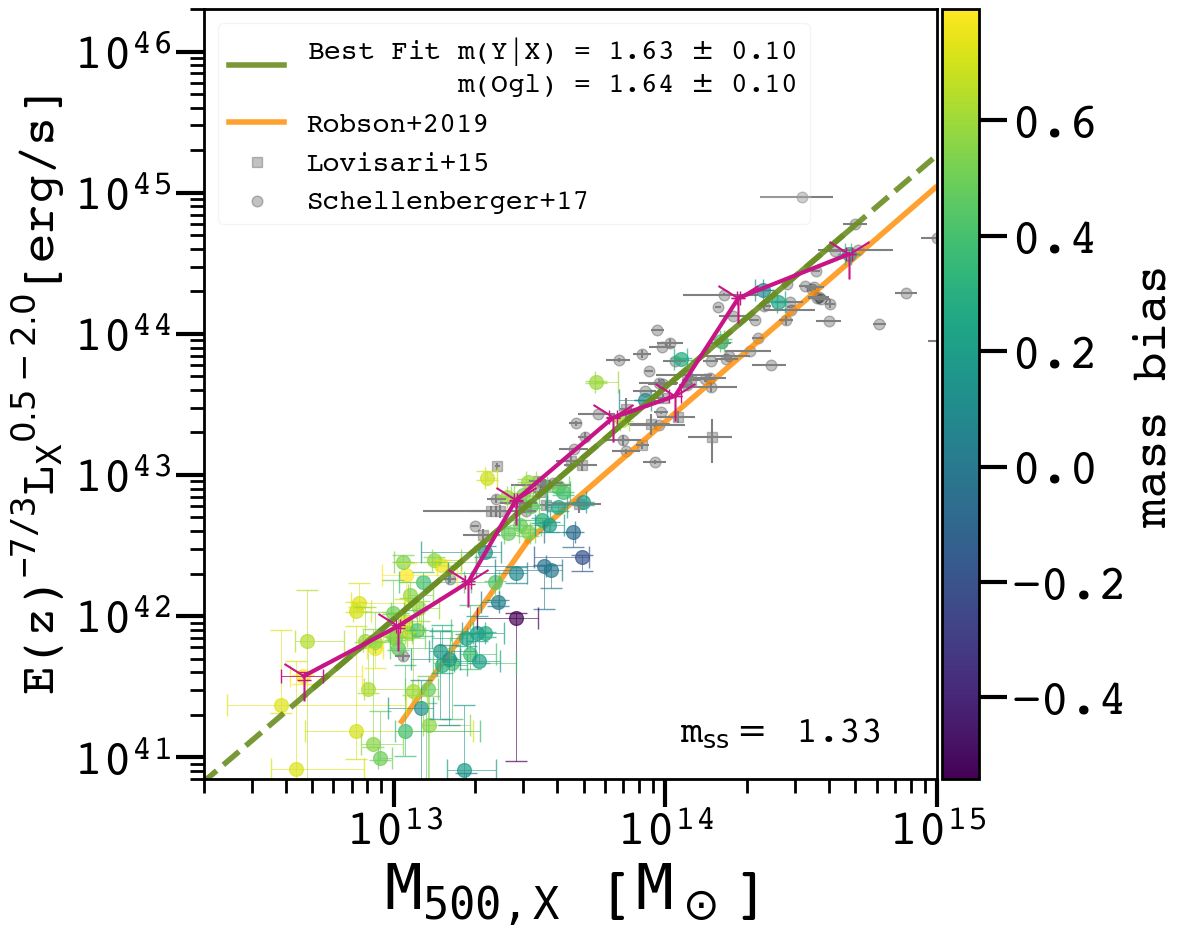} }}%
    \caption{The $L_\text{X}-M_{500}$ relationship is plotted for the true emission on the left-hand side and for the \moxha\ sample on the right. $L_\text{X}$ is measured between $0.5-2.0$ keV (rest-frame) in both cases, and within apertures of $R_\text{500,SO}$ and $R_\text{500,X}$ respectively. $L_\text{X}-M_{500}$ is suited to a double power-law fit. The \moxha\ sample has an increased normalisation due to the increased mass bias causing a net shift down the x-axis, which brings the sample data more in-line with the observational datapoints.}%
    \label{fig:L_vs_M}%
\end{figure*}

\begin{figure}
\centering
	\includegraphics[width=7.2cm]{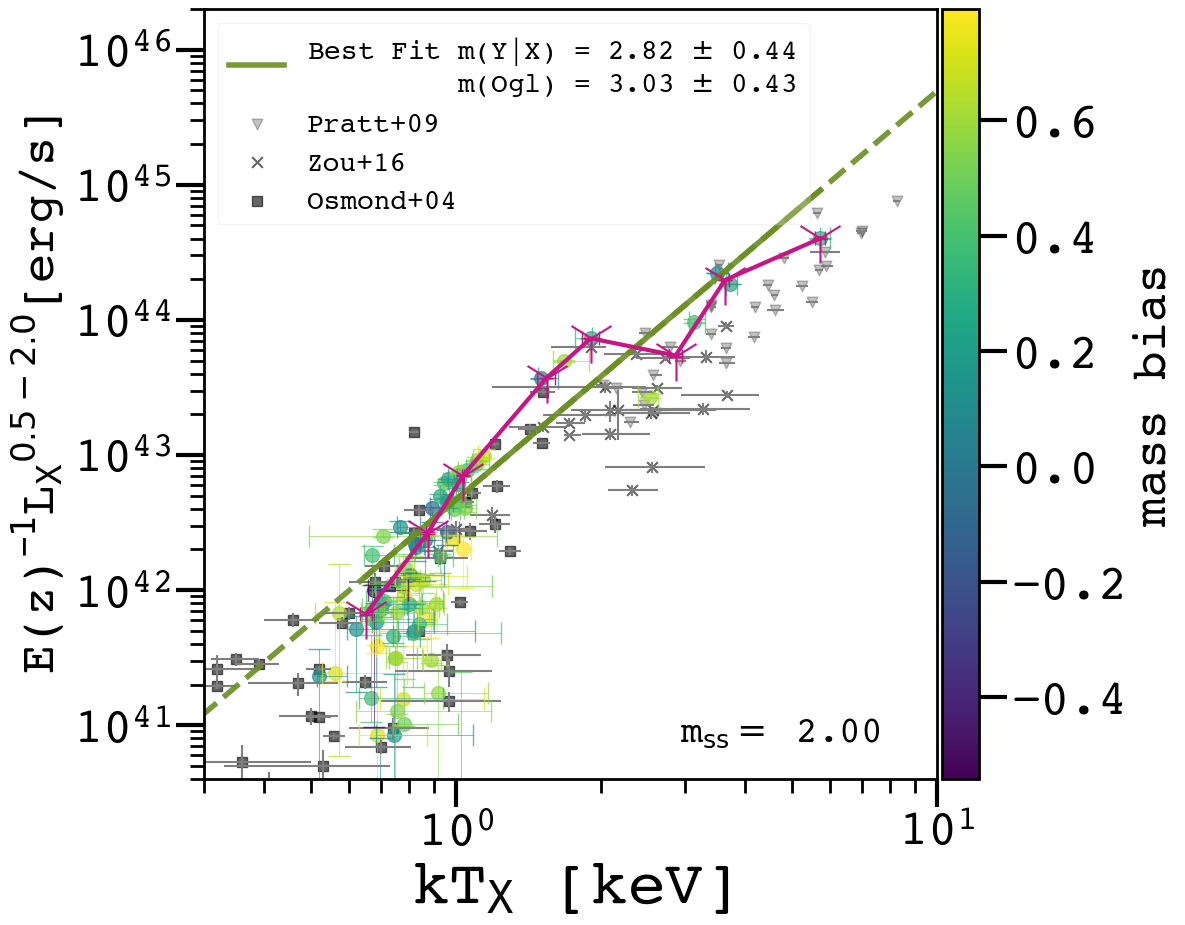}
    \caption{We plot the $L_\text{X}$ vs $kT_\text{X}$ relation for our \moxha\ sample. \simba halos reproduced the observed turn-off at $\sim 1$ keV, indicating that the feedback prescription is effective at matching the true baryonic content at the group scale.}
    \label{fig:L_vs_kT}
\end{figure}

\begin{figure*}%
    \centering
    \subfloat[\centering Mass-weighted]{{\includegraphics[width=6.8cm]{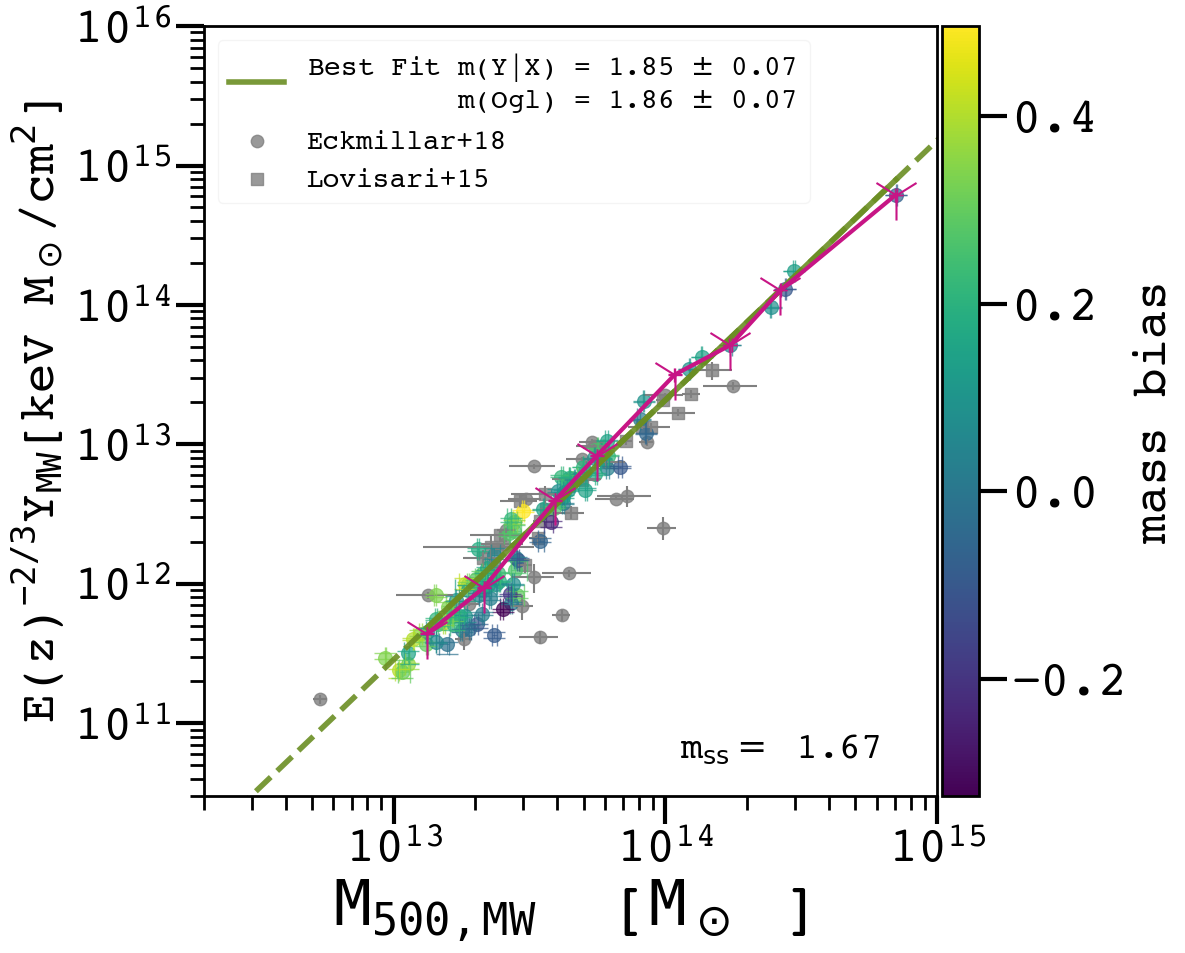} }}%
    \qquad
    \subfloat[\centering \moxha]{{\includegraphics[width=6.8cm]{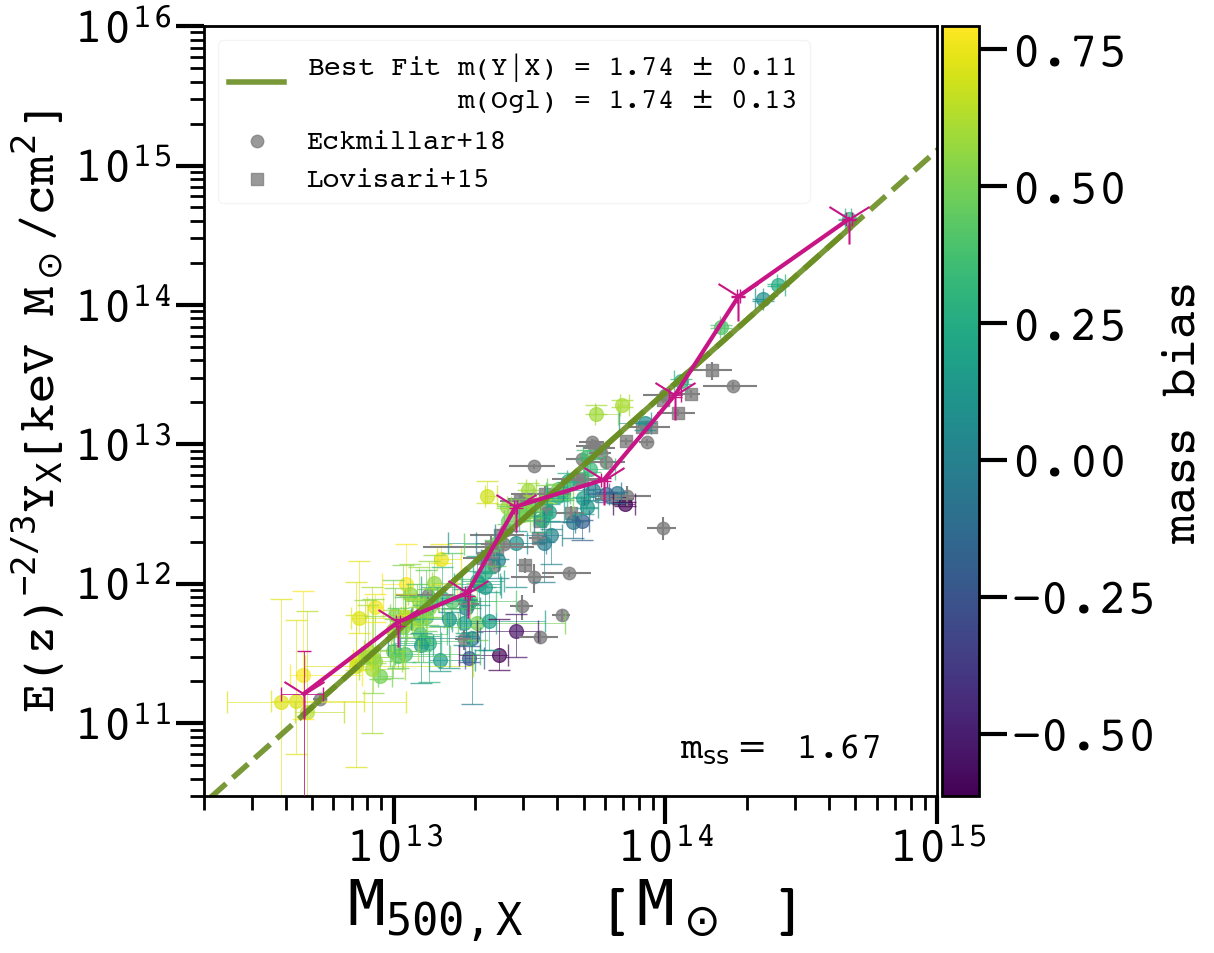} }}%
    \caption{We plot the $Y_\text{X}-M_{500}$ relation for our MW and \moxha\ samples. The best-fit slopes agree within error between the mass-weighted and \moxha\ samples. Both are close to the self-similar slope value but are steeper, matching the observational datapoints well.}%
    \label{fig:Y_vs_M}%
\end{figure*}

\begin{figure*}%
    \centering
    \subfloat[\centering Mass-weighted]{{\includegraphics[width=6.8cm]{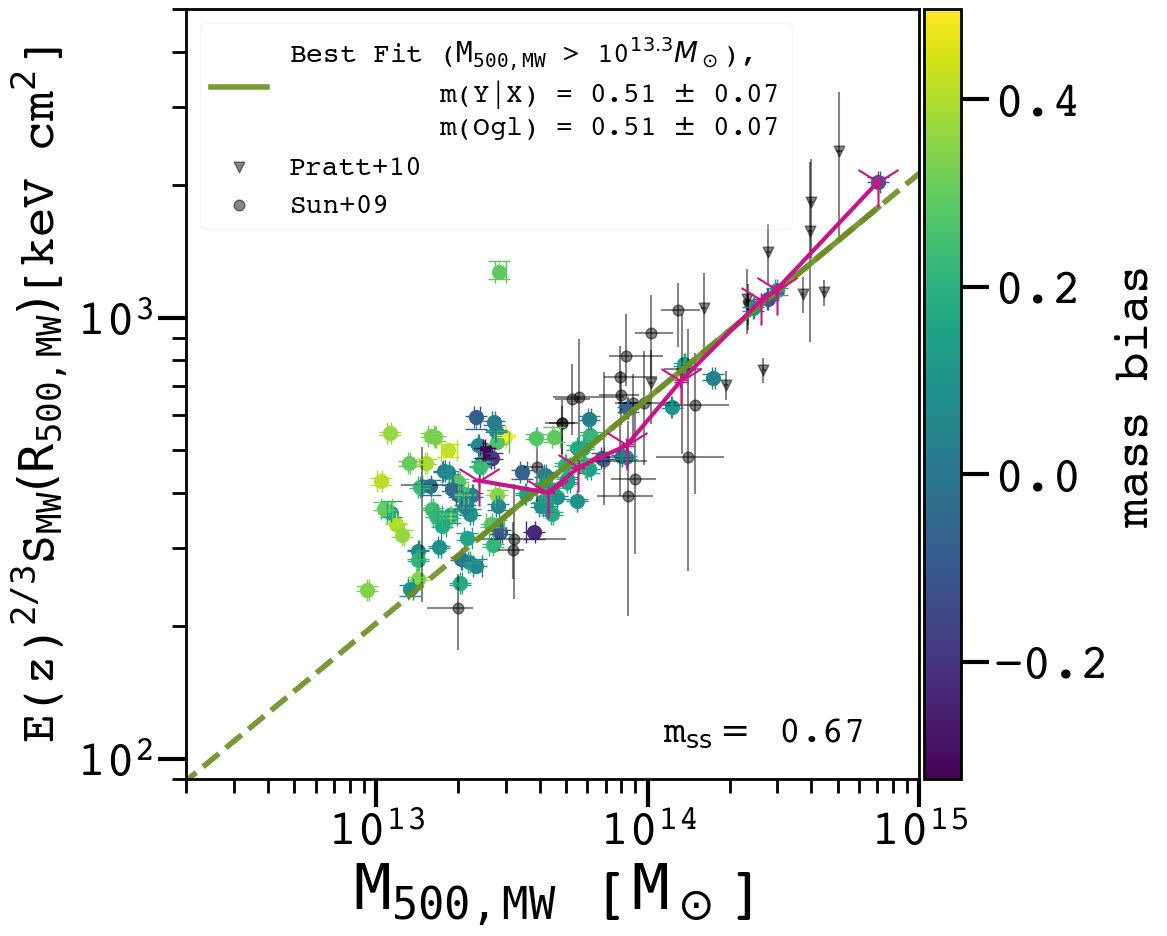} }}%
    \qquad
    \subfloat[\centering \moxha]{{\includegraphics[width=6.8cm]{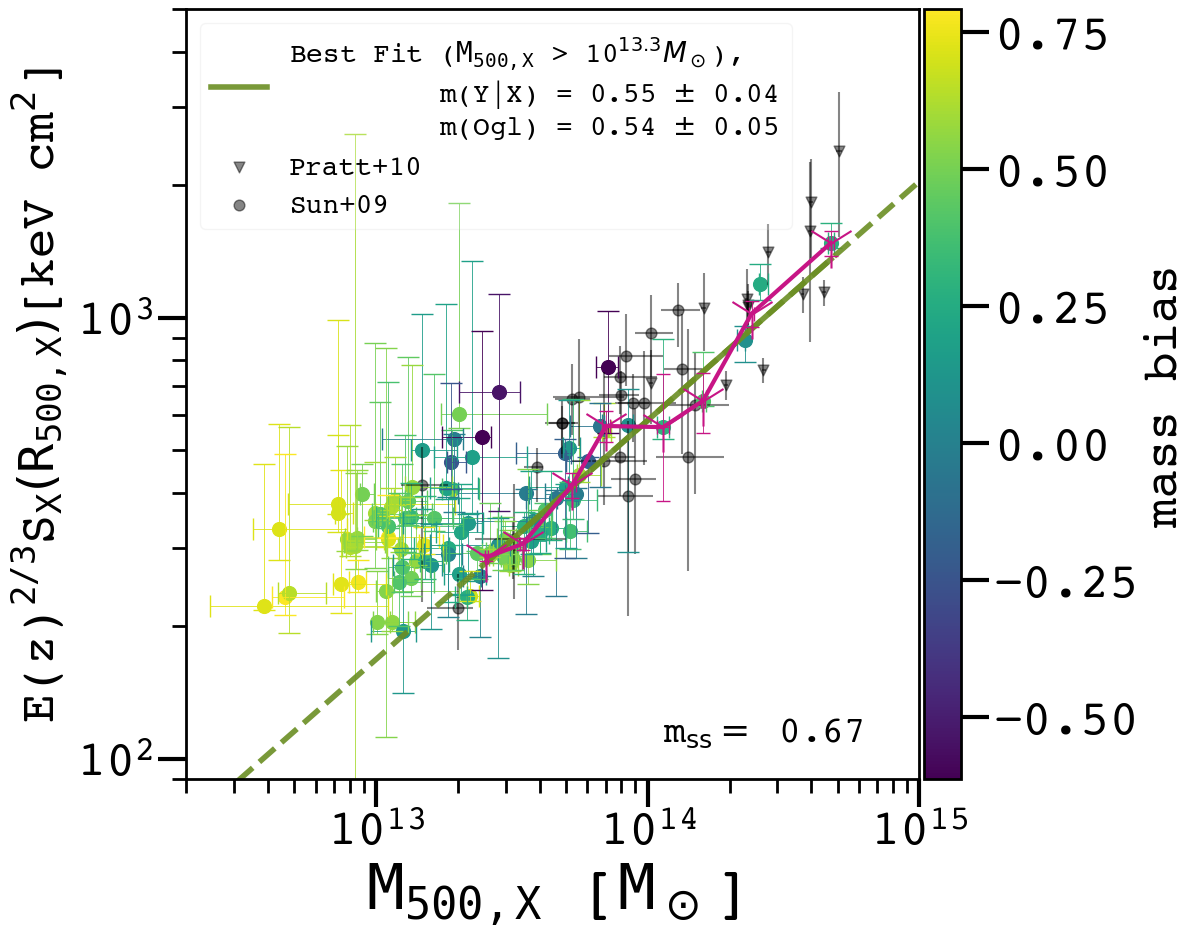} }}%
    \caption{We plot the $S(R_{500})-M_{500}$ relation for our MW and \moxha\ samples. Both MW and \moxha\ samples approach the observed slope from \citet{NagaiKravtsov2007} at high masses, but flatten significantly in the groups mass range. This indicates that the \ac{agn} feedback is able to drastically change the baryonic properties of low-mass halos out to large radii.}%
    \label{fig:S_vs_M}%
\end{figure*}

As another application of \moxha\ to the \simba simulations, we turn our attention to predicting the mocked X-ray scaling relations. We calculate the best-fit scalings for our mass-weighted and \moxha\ samples and compare to both the analytical predictions and to real observational data \footnote{We use reported individual redshifts to re-normalize the external datasets with the appropriate factors of $E(z)$. See \citet{LovisariMaughan2022} for a derivation of these factors in relation to the scaling relations. Where the dependence of an observational value on $h$ is provided alongside the observational data, we also re-normalise the observational data to our cosmology, with $h=0.68$.} Best-fit slopes are calculated using the bivariate correlated errors and intrinsic scatter (BCES) method for robust linear regression \citep{AkritasBershady1996}, taking into account (averaged plus/minus) errors on both the x-variable and y-variable. We use the \typefont{BCES} python package introduced in \citet{NemmenGeorganopoulos2012} to do this, and report values obtained with both the \textit{Y|X} (treats \textit{X} as the independent variable) and \textit{orthogonal} (minimises the orthogonal distance with no assumption on which variable is independent) methods.
To avoid a biasing of the scaling relations by the low end of the halo mass scale where we have many more objects, we bin the halos into (a maximum of) $8$ linearly-spaced logarithmic mass bins. We set the bin mass to the mean (log-)mass within the bin (to avoid a shifting of large mass bins where perhaps only one halo is present) and the bin $y-$value to the mean halo (log-)$y-$value within the bin. We determine the bin mean and errors on the mean $x$ and $y$ values, which are used for weightings in the linear regression, via 
\begin{equation}
\bar{x} = \frac{\sum x_i/\sigma_i^2}{\sum 1/\sigma_i^{2}}, \ \ \ \sigma^2_{bin} = \frac{1}{\sum 1/\sigma_i^{2}},
\end{equation}
where the $\sigma_i$ are (average $\pm$) individual halo uncertainties (in log space). It is these binned data points (shown in pink) which we then use to fit a scaling relation.

\subsection{Mass vs. X-ray Temperature}

Fig. \ref{fig:M_vs_kT} shows the $M_{500}-T_\text{X}$ relation computed using the mass-weighted hydrostatic mass and X-ray temperature (left panel) and the \moxha-inferred quantities (right panel).  Best-fit power-law relations to binned means (in pink) are shown as the green lines.  An assortment of observational measurements and fits are shown as black points and lines, as indicated in the legend.

Overall, the relations show a steep descent with a power law relatively close to the self-similar expectation of $M_{500}\propto T_\text{X}^{1.5}$.  In detail, for the MW case, a power law fit slightly steeper than self-similar is a good fit across all mass ranges, with a value of $1.92 \pm 0.06$ for an orthogonal BCES best fit over the entire mass range. 

Similar trends are seen in the \moxha\ halos (right panel). With all halos included in the fit, the slope is $1.72 \pm 0.15$ ($1.77\pm0.17$) for the $Y|X$ (Orthogonal) fit. There is an increased spread around the best-fit and running median at the low-mass end $\sim 1$ keV. By showing the measured mass bias on the colorscale, we see that the scatter below the trend is, as expected, dominated by those halos with a large mass bias. The same effect, although to a lesser extent, is displayed in the mass-weighted sample. We also see a systematic shifting of the halo temperatures to lower values for the \moxha\ sample compared to the mass-weighted sample. There is also a gap in the \moxha\ sample evident between $1-2$ keV; this may correspond to the temperature range in which the validity of a single-temperature fit begins to degrade \citep{MazzottaRasia2004,RasiaMazzotta2005}, which could also explain the larger scatter in the groups regime, along with the larger mass bias. Even with biasing in both mass and temperature (which of course are correlated) the \moxha\ sample remains fully consistent with observed data,  although the lack of good group samples at the very lowest masses makes it difficult to distinguish between the MW and \moxha\ cases in comparison to real data.

We so still compare the \simba results with the results of several observational catalogues. The \citet{VikhlininKravtsov2006} sample consists of $13$ relaxed low-redshift clusters with temperatures just below $1$ keV, observed with \textit{Chandra}. \citet{SunVoit2009} present $43$ \textit{Chandra} group samples, of which $23$ have mass estimates reported. \citet{EckmillerHudson2011} again use \textit{Chandra} data to present $26$ galaxy groups at low redshift, selected from a statistically-complete sample of $112$ X-ray selected groups. \citet{LovisariReiprich2015} provide data for $22$ \textit{XMM-Newton}-observed groups in a complete sample selected from the \textit{ROSAT} All-Sky Survey.

In all, our MW and \moxha\ results broadly agree well with the observational results, and with themselves, across the mass range where there is sufficient observational data. Below $1$ keV, where observations are thin, the \moxha\ sample predicts a larger scatter and slightly shallower best-fit scaling relation compared to the mass-weighted sample; this is due to an increased mass bias, and additional spectroscopic temperature bias, both of which are more severe at the low halo mass end.
Comparing the normalisation of the scaling relations obtained from \simba\ halos previously in \citet{RobsonDave2020} whose methodology is most similar to mass weighting, they noted that a $20-30\%$ bias in $M_{500}$ would coincide the simulated \simba\ halos with the groups of \citet{KravtsovVikhlinin2006} when comparing the $M_{500} - T_\text{X}$ relationship (not plotted here). Indeed we do see an improved agreement with the observational results having used hydrostatic mass estimates and spectroscopic temperatures, with our data agreeing better with the observational data when it comes to the normalisation of the relation. This is despite the fact that the mass bias is to first order due to a linear dependence on the normalisation of $kT(r)$, which we and others have found to be consistently biased low. 

We have further measured the scaling relations both for $kT_\text{X}$, weighted between $0.1-1R_{500}$, and $kT_\text{X}(R_{500})$. The best-fit slopes are similar, at $1.73 \pm 0.06$ and $1.68 \pm 0.28$ for the $Y|X$ fits for mass-weighted and \moxha samples respectively. The temperatures are naturally higher for the weighted quantity compared to the one measured at $R_{500}$, since the temperature profile of almost all halos drops with radius, with a few staying flat. Many comparative studies use a "global temperature" measure in $kT$ scaling relations, fitting a single spectrum gathered from a (core-excised) region from within $R_{500}$ (e.g. \citealp{LovisariReiprich2015}). We expect our weighted quantity in this case to be the best comparative value.

\subsection{Mass vs. X-ray Luminosity}

Our mock-observational $L_\text{X} - M_\text{500,X}$ and $L_{\rm truth} - M_\text{500,SO}$ relationships, plotted in Fig. \ref{fig:L_vs_M}, agree well with the broken power-law best-fit of \citet{RobsonDave2020}, although with a higher normalisation in the clusters range for the \moxha\ sample, and a lower normalisation in the groups range for the 'ground truth', intrinsic sample. The luminosity magnitudes for both samples are broadly similar (the maximum absolute difference in mean logarithmic value between the true luminosity and the X-ray luminosity bin one against the other is small, at $0.26$). We hence associate the apparent increased normalisation in the \moxha\ case compared to the intrinsic to the biasing of $M_{500}$ below the true virial mass, which is a larger effect at the low mass end. 

Along with the sample of \citet{LovisariReiprich2015}, we compare also to the $64$ \textit{Chandra} clusters that make up the HICOSMO catalogue of \citet{SchellenbergerReiprich2017} (we use their "NFW-Freeze" model best-fit values). The \moxha\ data still fits well within the observed luminosities of \citet{SchellenbergerReiprich2017} and \citet{LovisariReiprich2015}, especially for the clusters. Interestingly, we find that the scatter in the luminosity is much better reproduced by the fully mocked sample than for the intrinsic one. At the high-mass end both intrinsic and X-ray samples produce good agreement with observed results, though drawing conclusions on the scatter is difficult because of the limited sample size there.

Looking at the best-fit scaling relations, it is clear that a single power law fit to both the intrinsic data does not capture the mass-dependence of the luminosity very well, with a broken power law probably a better match. The feedback in \simba\ is able to reduce the amount of luminous gas in these group objects, and this is clear in the intrinsic steepening of this scaling relation. It cannot be a selection effect, since this would result in an artificial \textit{over}-brightening of groups.  This steepening is too severe compared to the observed data we compare to, indicating that the energy of \simba\ feedback coupling to the \ac{igrm} is too large a fraction of the halo binding energy at the low mass end.

For the \moxha\ data, we again find mostly good agreement with the observational data, and find better agreement in the magnitudes of the luminosities than was found with the intrinsic sample, where we found that the group-scale objects are potentially too dim. The strong effect of removal of cold gas on the luminosity scaling relation seems to bring \simba\ groups well in-line with observational data, despite the lower baryon fraction and gas density radial profiles compared to other simulations \citep{OppenheimerBabul2021}. 

Again, we see that a mass bias at the low-mass halo end has the effect of shallowing the observed scaling relation, partially masking the true steepening compared to self-similarity.

In the \moxha\ data the large scatter at the low mass end makes deciding between a single or double power law model more difficult. Because the uncertainties in the luminosity are calculated through repeating the luminosity measurement within different apertures given by the confidence interval on $R_\text{500,X}$, the $x$ and $y$ errors in this case are correlated, and become large at the low-mass end.

\subsection{Luminosity vs. Temperature}

We now turn attention to the $L_\text{X} -T_\text{X}$ relation as measured in our \simba\ sample by \moxha. We plot this in Fig. \ref{fig:L_vs_kT}. Again we have a difficulty in extracting good signal to noise for the very lowest mass groups, as seen in the large vertical error bars at the low-mass groups end, which is something we encountered above in the $L_\text{X}-M_\text{500,X}$ scaling relationship.

The sharp turn-off that is clearly seen at $\sim 1$ keV in Fig. \ref{fig:L_vs_kT} makes fitting a single power law to halos below this temperature (mass) scale difficult, and indeed we find poor fits to the data when binned over the whole mass range with both BCES methods with large uncertainties. Fitting to the whole sample, we consistently find a higher-than-self-similar value in both the weighted $kT_\text{X}$ and $kT_\text{X}(R_\text{500,X}$ values. This value is $2.82 \pm 0.44$ ($3.03 \pm 0.43$) for the BCES $Y|X$ (orthogonal) best-fit to the weighted temperatures, and $2.93 \pm 0.48$ ($3.12 \pm 0.47$) for the scaling relation of $kT_\text{X}(R_\text{500,X})$ (not plotted). These values are high compared to the self-similar prediction of $2$, but very much in-line with observed values \citep{LovisariEttori2021}.

The fact that the temperature value \textit{at} $R_\text{500,X}$, $kT_\text{X}(R_\text{500,X})$, demonstrates a similar steepening compared to self-similar as the weighted temperature value again demonstrates how \simba\ feedback is able to drastically change gas properties at large radii from the \ac{bcg}/\ac{bgg} itself, as we saw previously in our measured $M_{500}-T_\text{X}$ relations.

To compare to real halos, we also plot results from several observational studies. Firstly, $31$ clusters from the Representative XMM-Newton Cluster Structure Survey (REXCESS) as presented in \citet{PrattCroston2009}, which have temperatures from $2-9$ keV. At lower masses we plot the group and low-mass cluster data of \citet{ZouMaughan2016}, and also the $45$ \textit{ROSAT}-observed groups of \citet{OsmondPonman2004} which have spectroscopic data \footnote{Redshifts are not provided in this study, so we use a constant $z = 0.1$ to re-normalize the data points in the scaling relation.}.

Our observed scatter agrees well at the high temperature end with the observational data of \citet{PrattCroston2009} and \citet{ZouMaughan2016}, although possibly the clusters are slightly too bright or slightly too cold. It is at the low mass end where we find the most interesting comparison. Our observed turn-off at $\sim 1$ keV is able to very closely reproduce the turn-off observed in the sample of \citet{OsmondPonman2004}. This indicates that the \simba\ feedback model is broadly successful, at least in this region of $L_\text{X}-T_\text{X}$ space, in evacuating sufficient cold fluid from groups to reproduce the observed sharp turn-off in temperature and luminosity relations. This appears to happen at the correct temperature scale and with the required steepness. However, we do not reproduce the coldest groups observed in the aforementioned observations. This could again be indicative that the \simba\ feedback model is too strong, and though it reproduces the low-luminosity/over-heated turn-off observed in both the $L_\text{X}-M_{500}$ and $L_\text{X}-T_\text{X}$ relations, it cannot reproduce the observed population of less affected clusters which retain cold, low-entropy, luminous gas within $R_{500}$, at least in the sample tested in this work.

\subsection{Integrated Compton $Y$ vs. Mass}

We plot our mock-observed $Y_\text{X} - M$ scaling relation in Fig. \ref{fig:Y_vs_M}. $Y_\text{X}$ is a low-scatter cluster/group-mass proxy introduced in \citet{KravtsovVikhlinin2006}, insensitive to the relaxed-ness of the system. This parameter is robust and low-scatter because the biases in $M_{500,gas}$ and $T_\text{X}$ anti-correlate. 
And unsurprisingly indeed we do find the lowest apparent scatter out of all relations in $Y- M_{500}$, both for the mass-weighted sample in the left hand panel of Fig. \ref{fig:Y_vs_M} and in the fully-mocked X-ray \moxha\ sample on the right hand side. 
For the mass-weighted sample we find the slope to be $1.85 \pm 0.07$ ($1.86 \pm 0.07$) and for \moxha\ it is $1.74 \pm 0.11$ ($1.74 \pm 0.13$). Both are therefore steeper than the self-similar prediction of $5/3$ which agrees with the findings of \citet{PopHernquist2022}, who predict a steepening of $Y_\text{X}-M_{500}$ due to groups. The increased mass bias for the lowest mass groups is again apparent as a driver of the shallowing of the scaling relation, with a clear stratification in average bias observed above and below the best-fit trend. As expected however, the spread is less severe than for other scaling relations and the \moxha\ best-fit slope agrees with MW within error.

In all, our \moxha\ data matches the mass-weighted sample as well as observational data very well, although again the lack of good group samples at the very lowest masses limits good comparison to simulated values and scatter. \\

\subsection{Mass vs. Entropy}
Finally, in Fig. \ref{fig:S_vs_M} we plot the $S_\text{X}$ vs $M_\text{500,X}$ scaling relation. We fit core-excised radial profiles in this work, with only radii $\geq 0.1R_\text{500,SO}$ being considered, so we do not plot $S(0.1R_{500})$ as is commonly done in similar works. Instead we opt to plot the quantity at the hydrostatic estimate of $R_{500}$. We plot for comparison the observational results of \citet{SunVoit2009} and \citet{PrattArnaud2010}, both of which use a similar definition of the entropy in their quoted values.

In the high-mass group and cluster regime, \simba\ matches the self-similar prediction fairly well in both the mass-weighted and \moxha\ samples. For instance, the best-fit for the mass-weighted sample is $0.51 \pm 0.07$ ($0.51 \pm 0.07$) for $M_\text{500,MW} > 10^{13.3}M_\odot$, and the \moxha\ best-fit slope is is $0.55 \pm 0.04$ ($0.54 \pm 0.05$) for $M_\text{500,X} > 10^{13.3}M_\odot$, not far from the self-similar value of $2/3$. However, all entropy scalings we find become shallower once groups are included, and flatten out almost fully deep into the groups regime for the \moxha\ sample. This is another evidence showing how \simba's feedback model is effective in modifying the halo gas properties out to large radii in groups. Cold, low entropy gas at and near the virial radius in groups is heated and swept up in feedback from the central source, causing a flattened $S_\text{X}-M_{500}$ relation at the low mass regime in both weighted and \moxha\ samples. There is also increased scatter here, because the feedback does not impact all groups as equally as gravitational effects do, and stochastic, directional jet events interact in a more complex manner with the cool fluid.

Both sets of observational data match very nicely with both the MW and \moxha\ sample entropies. The \moxha\ sample does seem to better agree with both the scatter and uncertainties reported for observed halo, although observational data is scarce in the low-mass groups range where the scatter is largest.

\subsection{Mass Dependence of the Scaling Relations}
\begin{figure}
	\includegraphics[width=\columnwidth]{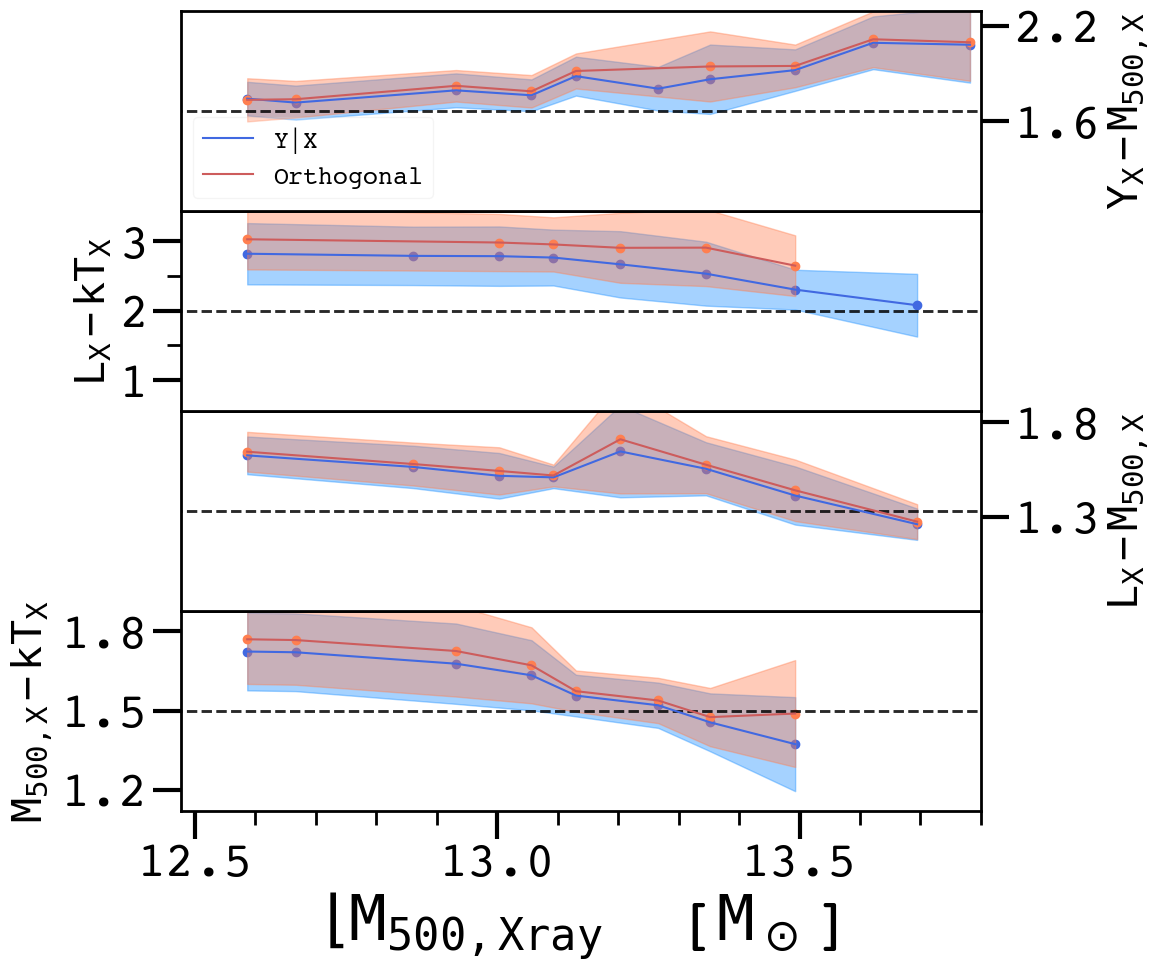}
    \caption{We plot the dependence of scaling relation best-fit slope on the minimum halo mass used in the fitting, for our fully mocked \moxha\ sample. We plot all points which are fit by the BCES method with reasonable error. Reading $\lfloor M_{500}$ from high to low, each point includes $10$ more halos in the fit than the last. The first point includes the $10$ highest-mass halos, and the final point includes all halos in the sample.}
    \label{fig:scalings_vs_Mmin_panel_Xray}
\end{figure}

Finally we investigate the effect the minimum mass of the halo sample has on the scaling relation slope. We are interested in quantifying how the low-mass groups differ from the clusters, and how limited observational samples only including halos above a given mass (alternatively luminosity) may measure a different scaling relation than a more complete one. This problem is especially potent when one desires a scaling relation slope, e.g. $Y_\text{X}-M_{500}$, in order to estimate halo masses from a proxy. 

Fig. \ref{fig:scalings_vs_Mmin_panel_Xray} shows this dependence of slope on minimum halo mass, for various scaling relations.  To investigate this mass dependence we calculate the BCES line of best-fit using both the \textit{Y|X} (shown in \textit{blue}) and \textit{orthogonal} (\textit{red}) methods for binned data in the same manner we used for the scaling relations. We re-iterate that we bin the data prior to fitting in order to avoid biasing the fit due to the larger number of low-mass halos in our sample. We start by setting the minimum mass to be the $10$th largest mass in the sample, calculate the best-fit, then repeat, each time adding in the next lowest mass halo (with mass we denote $\lfloor {M}_\text{500,X}$) until we have reached the halo with the lowest mass of our sample and as such all halos are included in the fit. We then plot the best-fit scaling slope against $\lfloor {M}_\text{500,X}$ in Fig. \ref{fig:scalings_vs_Mmin_panel_Xray}. The confidence interval shown is the spread on each data point as calculated by BCES through $10,000$ bootstrapping trials. 

It is clear that the inclusion of groups in the fully mocked sample tends to steepen the $M_{500}-T_\text{X}$ scaling relationship (bottom panel), indicating an over-heating of halos at this low-mass end. This also is seen when the same analysis is performed for the mass-weighted quantities (not plotted). 

The luminosity relations (middle two panels) only tend to approach self-similar when the \moxha\ sample is reduced to the highest-mass objects, and tend to asymptote towards a steeper than self-similar slope with the inclusion of more and more low mass halos, and the slope tends to gradually steepen when lower mass halos are included, demonstrating the under-luminous nature of groups relative to clusters due to more efficient expulsion of cold gas. 

Perhaps of most interest is the trending of the $Y_\text{X}-M_{500}$ relation (top panel) in the \moxha\ quantities to the theoretical slope value at low halo masses, indicating the importance of fully exploring the groups regime when obtaining $Y_\text{X}$ to be used as a mass proxy. In the mass-weighted quantities, the inclusion of groups in the fits tends to instead slightly steepen the $Y_\text{X}-M_{500}$ relation, better matching the results of e.g. \citet{PopHernquist2022}.

Overall, the scaling relations are strongly dependent on the minimum halo mass in the sample, with poorer groups being under-luminous and over-heated relative to more massive groups. This trend is also clear in the MW sample too. We finally note that the \textit{orthogonal} BCES method tends to systematically obtain a steeper best-fit line that \textit{Y|X} in all cases.

\section{Summary}\label{sec:Discussion}

In this work we have introduced a new end-to-end X-ray pipeline, \moxha, building upon the emission calculations of \PyXSIM, detection facilities of \SOXS, and spectral-fitting capabilities of \texttt{threeML}, to generate measurements of the X-ray mass bias and halo scaling relations for halos within the \simba\ full-physics cosmological simulation. Included are additional routines concerning annulus generation, deprojection, substructure masking, background subtraction, and profile fitting. We use \moxha\ to perform an in-depth study of the largest $\sim 100$ low-redshift halos from \simba that are sufficiently X-ray bright to reliably X-ray spectral fit in order to extract physical quantities. We have several main findings:
\begin{itemize}

    \item We report a hydrostatic mass bias of $b_\text{MW} = 0.15 \pm 0.019$ underestimating the true $M_{500}$ in \simba when considering mass-weighted density and temperature profiles. There is a systematically higher mass bias of $b_\text{X} = 0.41 \pm 0.03$ from fully-mocked, \moxha\ quantities. This bias is mildly reduced to $b_{X,ODR} = 0.33 \pm 0.04$ when the profiles are fit with an ODR method. The \moxha  biases are roughly consistent with but slightly larger than the bias inferred from emissivity-weighting and luminosity-weighting: $b_\text{EW} = 0.36 \pm 0.02$ and $b_\text{LW} = 0.30 \pm 0.02$ respectively. All biases show only a weak dependence on halo mass, with the \moxha-inferred bias becoming mildly larger below $M_{500}\la 10^{13.5}M_\odot$.
    
    \item By comparing across the various measured bias values, we quantify the bias contribution from observational effects such as projection effects, background and instrumental noise, and foreground absorption. We find that the largest increase in bias is associated with the move from mass-weighted to emission-weighted quantities, constituting an increase in $b$ by around a factor $\times 2$, with the absolute scatter remaining roughly the same. Going from emission-weighting to full \moxha (i.e. including observational effects) incurs an additional maximum median bias of $\sim +~0.1$, and notably increases the tail of the bias distribution towards low-biases.

    \item Through the two-sample K-S test we find statistical agreement between the probability distributions of the emissivity-weighted bias with both the luminosity-weighted counterpart, as well as the \moxha-inferred bias distribution, at the $95\%$ confidence level. The K-S statistic for K-S$(b_\text{EW},b_\text{X})$ and K-S$(b_\text{EW},b_\text{LW})$ were $0.17$ and $0.086$ respectively, which are statistically not clearly distinguished. In contrast, we find a significant difference of the mass-weighted bias versus all other bias measures. 
    
    \item We further considered a shifted two-sample K-S test in which we test the distributions' similarities once they have been shifted to the same mean.  We find that the mass-weighted bias is not distinguishable from the others in the shifted K-S test, although the consistency is much stronger for the luminosity and emissivity weighted biases as opposed to the \moxha-inferred bias.  This indicates that once the overall bias shift has been factored out, the range of biases among the various measures are concordant.

    \item The mass-weighted $M_{500}-T_\text{X}$ relation appears fairly well-fit by a steeper-than-self similar slope of $1.92 \pm 0.07$, while the \moxha\ sample shows a corresponding slope of $1.72\pm0.15$. This trend is similar when the temperature is instead measured at $R_{500}$. \simba\ feedback is sufficiently strong as to efficiently heat groups to a larger degree than clusters, leading to a steeper than self-similar scaling relation. This heating is effective out to a large radius, changing the gas properties at $R_{500}$ or beyond. Increased temperature bias below $\sim 1$ keV leads to an artificial shallower slope in the \moxha\ sample compared to mass-weighted, which partially masks the true steepening of the relationship.

    \item The $L_\text{X}-M_{500}$ scatter for the intrinsic and \moxha fits the results of \cite{RobsonDave2020}, who use a broken power law with pivot at $3 \times 10^{13} M_\odot$ to fit the \simba\ halos. The \moxha\ sample has a slightly higher normalisation because of the mass bias, compared to the mass-weighted scatter. The increased scatter in the X-ray relation makes accurate determination of the intrinsic scaling relations difficult at the group scale, and the case for a broken power-law best fit is less strong. That said, the \moxha sample better matches the observed spread in $L_\text{X}-M_{500}$ from \citet{SchellenbergerReiprich2017} and \citet{LovisariReiprich2015}, versus that of the intrinsic data.

    \item The $L_\text{X}-T_\text{X}$ scaling has a sharp turn-off at $\sim 1$ keV for our X-ray sample, excellently matching observed halos of \citet{OsmondPonman2004}. This  indicates that \simba\ feedback is effective in removing cold gas from groups and reproducing the overheating of such halos as observed in real data. \simba however lacks the population of cold groups that is also observed, suggesting that the \simba\ feedback prescription may be too ubiquitous.

    \item The scatter in our observed scaling of the robust mass proxy $Y_\text{X}$ vs $M_\text{500,X}$ and $Y_\text{MW}$ with $M_\text{500,MW}$ is low. The mass-weighted best fit is $1.85 \pm 0.07$, and for the \moxha\ sample is $1.74\pm 0.11$. The MW sample agrees with the trend of \citet{PopHernquist2022} from IllustrisTNG, where there is a steepening in the relation in the groups regime; this steepening is washed out by larger mass bias for the lowest mass objects in the \moxha\ sample. Both of our samples well match the observed data of \citet{EckmillerHudson2011} and \citet{LovisariReiprich2015}. The lack of observational data for groups with the same quality and sample size as for clusters makes robust comparison in the low-mass end difficult.

    \item The \simba\ halos match the observed $S_\text{X}-M_\text{500,X}$ scatter of \citet{SunVoit2009} and \citet{PrattArnaud2010}, with the spread on the \moxha sample closer to the observational data. At low group masses this relation is flat, with the slope at higher masses close to the self-similar value.

    \item By plotting the scaling relations obtained as a function of the minimum halo mass in the sample, we have shown that the scaling relations are fairly sensitive to the sample's halo mass range and only seem to converge near to the self-similar slope values at the highest mass halos in \simba. Including poorer groups is clearly seen to steepen the $M_{500}-T_\text{X}$ and $L_\text{X}-T_\text{X}$ relations, owing to the increased relative temperatures and lowered baryon contents in $\la 10^{13.5}M_\odot$ systems putatively driven by \simba's AGN feedback.

\end{itemize}
This work serves as both as an example of using the \moxha pipeline for measuring mass bias and X-ray scaling relations, and also as a basis for more in-depth studies of halos in \simba and other cosmological simulations with \moxha. Accurate mock observations are essential for robustly testing the increasingly detailed physics in modern simulations against upcoming X-ray data. This data will come from one or more planned new X-ray missions, for example: {\it Athena}, which we used in this work (see description in \hyperref[sec:Generating Mock Observations]{\S3}); {\it LEM} \citep{KraftMarkevitch2022} which is a proposed high spectral-resolution mission with a large grasp, designed to probe faint extended objects, including the imprints of \ac{agn} feedback in groups and clusters \citep{SchellenbergerBogdan2023}; and {\it XRISM} \citep{XRISMScienceTeam2020}, which also will perform high-resolution spectroscopy of groups and clusters, including a focus on gas kinematics. \moxha represents a step forward in enabling such analyses for a wide range of simulations that are compatible with \yT, providing the community with a valuable tool for exploring the physics associated with X-ray emitting gas in the universe.

\section*{Acknowledgements}
FJ would like to acknowledge the support of the Science and Technology Facilities Council. FJ would like to thank Arif Babul, Benjamin Oppenheimer, and Adam Ormondroyd for useful conversations, and Douglas Rennehan for useful conversations and comments on the manuscript. He would also like to thank Britton Smith for advice regarding \yT\, and John ZuHone for advice regarding the \PyXSIM and \SOXS software packages, as well as for making these packages publicly available. Both authors would like to thank the anonymous referee for a constructive report which improved the content and format of the work. For the purpose of open access, the author has applied a Creative Commons Attribution (CC BY) licence to any Author Accepted Manuscript version arising from this submission. 

\section*{Data Availability}

The \simba simulation data used in this work is publicly available at {\tt https://simba.roe.ac.uk}.
The \moxha\ package will be made available and open-access  in due course, while for now it will be shared upon reasonable request to the corresponding author.



\bibliographystyle{mnras}
\bibliography{example} 




\appendix


\bsp	
\label{lastpage}
\end{document}